\RequirePackage{fix-cm}

\documentclass[twocolumn,epjc3,amssymb,eqsecnum,amsfonts,epsf,preprintnumbers,nofootinbib,superscriptaddress]{svjour3}
\smartqed
\RequirePackage{graphicx}
\usepackage{multirow}
\usepackage{amssymb,graphics}
\usepackage{graphics,epsfig,subfigure}
\usepackage{epstopdf}
\usepackage{graphicx}
\usepackage{dcolumn,color}
\usepackage{bm}
\usepackage{epsfig}
\usepackage{hyperref}
\usepackage{color,framed}
\usepackage{lipsum}
\usepackage{amsmath,amssymb,graphics,epsfig,subfigure,amsfonts,latexsym,color,makeidx}
\usepackage{cite}

\journalname{Eur. Phys. J. C}

\begin{document}
\renewcommand{\baselinestretch}{1.3}
\title{Classification of Gravitational Waves in Higher-dimensional Space-time and Possibility of Observation}

\author{Yu-Qiang Liu\thanksref{addr1,addr2}
\and  Yu-Qi Dong\thanksref{addr1,addr2}
\and Yu-Xiao Liu \thanksref{e1,addr1,addr2}}

\thankstext{e1}{e-mail:liuyx@lzu.edu.cn, corresponding author}

\institute{Lanzhou Center for Theoretical Physics, Key Laboratory of Theoretical Physics of Gansu Province, School of Physical Science and Technology, Lanzhou University, Lanzhou 730000, China\label{addr1}
\and
Institute of Theoretical Physics and Research Center of Gravitation, Lanzhou University, Lanzhou 730000, China \label{addr2}}


\maketitle
\begin{abstract}
The direct detection of gravitational waves opens the possibility to test general relativity and its alternatives in the strong field regime. Here we focus on the test of the existence of extra dimensions. The classification of gravitational waves in metric gravity theories according to their polarizations in higher-dimensional space-time and the possible observation of these polarizations in 3-dimensional subspace are discussed in this work. And we show that the difference in the response of gravitational waves in detectors with and without extra dimensions can serve as evidence for the existence of extra dimensions.
\end{abstract}
\section{Introduction}
Since the gravitational waves of the collision of two black holes was first observed by Advanced Laser Interferometer Gravitational-wave Observatory (aLIGO) and Virgo collaborations in 2015 \cite{abbott2016observation,abbott2016gw151226}, we have entered the era of gravitational waves astronomy. So far,  over a hundred gravitational wave events have been detected by LIGO and Virgo \cite{abbott2021gwtc}. Among these events, some mergers of binary neutron stars were also observed. The most famous one is GW170817 \cite{abbott2017gw170817}, which is the first event with observable electromagnetic signal. All these events, especially for multi-messenger events like GW170817 \cite{creminelli2017dark,sakstein2017implications,gong2018constraints}, provide the test of gravity theories in strong gravitational field regime \cite{de2016constraining,scientific2016tests}.
\par
Einstein's general relativity has withstood all these most strict tests so far. It is still the most successful gravity theory. However, since the deepest disagreement between  general relativity and quantum field theory \cite{goroff1986ultraviolet,t1993one}, as well as the problem of dark matter \cite{zwicky1937masses} and dark energy \cite{peebles2003cosmological} etc., general relativity is generally regarded as just an effective theory. Therefore some modifications to it must be considered \cite{clifton2012modified}. These modifications can be roughly divided into three categories: higher dimensions, e.g. Kaluza-Klein (KK) theory \cite{kaluza1921unity}, domain wall model \cite{rubakov1983we}, large extra dimensional brane model \cite{arkani1998hierarchy}, warped extra dimensional brane models \cite{randall1999large,randall1999alternative}; violation of Lorentz invariance, e.g. Einstein-aether theory \cite{jacobson2004einstein}; more degrees of freedom, e.g. scalar-tensor theory (Brans-Dicke theory \cite{brans1961mach}, Horndeski theory \cite{horndeski1974second}), tensor-vector-scalar theory (TeVeS) \cite{bekenstein2004relativistic}. All these researches have fruitful results.
Most of these modified gravity theories have been constrained much more strictly after the direct detection of gravitational waves. Especially for the multi-messenger events like GW170817, the strongest limit to date on the speed of gravitational waves has ruled out some modified theories, e.g. quartic Galileons theory, de Sitter Horndeski theory etc. \cite{ezquiaga2017dark}. In this work, we only concern the higher-dimensional metric gravity theories.
\par
It is worth noting that the luminosity distance between the sources and observer can be obtained by the directly detecting the gravitational waves \cite{holz2005using,Deffayet:2007kf,Khlopunov:2022jaw,Khlopunov:2022ubp}. The luminosity distance depends only on the  ratio of the radiation density near the source to that near the observer if the space-time dimension is detemined.
 
  Therefore, if gravitational waves and electromagnetic waves from the same source all propagate in 4-dimensional space-time, their luminosity distance should be equal. But if there are extra dimensions, the luminosity distances calculated by gravitational waves and electromagnetic waves should different, since gravitational waves propagate in higher-dimensional space-time, while electromagnetic waves propagate only on a 3-dimensional subspace \cite{pardo2018limits,corman2021constraining}. At present, there is only one confirmed multi-messenger gravitational wave event, that is GW170817. The corresponding gravitational wave luminosity distance is $40_{-14}^{+8}$ Mpc \cite{abbott2017gw170817} and the electromagnetic wave luminosity distance is $40.7_{-2.4}^{+2.4}$ Mpc \cite{Cantiello:2018ffy}. So the possibility that the two luminosity distances are not equal can not be ruled out now. It is worth noting that the famous Dvali-Gabadadze-Porrati (DGP) model \cite{dvali20004d} theoretically fulfills the assumption that the extra dimensions occur only at large scale. Therefore, the difference between the two luminosity distances is allowed without violating current observations (there is another interesting model \cite{Mureika:2011bv, Stojkovic:2013xcj, Dai:2014roa} which assume that the dimension is low than 4 at short scale, and this will also lead to some interesting observable effect on gravitational waves). This is one of the motivations for studying gravitational waves in higher-dimensional space-time, and this is also one of the possible ways to detect the existence of extra dimensions. \par
Besides that, the difference in arriving times of gravitational signal and the corresponding light signal come from the same source has also been found in GW170817 \cite{abbott2017gw170817}. In extra dimensional theory, all interactions except gravity are trapped in a 4-dimensional brane \cite{ishihara2001causality,caldwell2001shortcuts,yu2017probing,lin2020constraint}. If the brane is not flat, then from the perspective of higher-dimensional space-time, electromagnetic waves will propagate along the brane, while gravitational waves can propagate in the whole higher-dimensional space-time \cite{yu2017probing,lin2020constraint,Visinelli:2017bny}. This may cause gravitational waves from the same source to reach the detector before electromagnetic waves. Therefore, from the perspective of 4-dimensional space-time, the propagation speed of gravitational waves may be faster than that of light, although the propagation speed of gravitational waves is still the speed of light in the whole higher-dimensional space-time. This is also one of the possible ways to detect the existence of extra dimensions. In this work we focus on another possible way, which is the polarization modes of gravitational waves.
\par
General relativity predicts two tensor polarization modes of gravitational waves, namely plus mode ($+$) and cross mode ($\times$). When generalize general relativity to most general metric theories in 4-dimensional space-time, there are at most six independent polarization modes, besides two tensor modes ($+$ and $\times$) there are two vector modes ($x$), ($y$) and two scalar modes: breathing mode ($b$) and longitudinal mode ($l$) \cite{eardley1973gravitational}.  Many works have been done on this topic, the possible polarization modes in the specific 4-dimensional modified theories such that $f(R)$ theory \cite{gong2018polarizations}, Brans-Dicke theory \cite{capozziello2006scalar,maggiore2000detection}, Horndeski theory \cite{hou2018polarizations}, Palatini Horndeski theory \cite{dong2021polarization}, massive gravity theory \cite{hyun2019exact} have been calculated. There are also some works on the polarization modes of gravitational waves in the theory of higher-dimensional space-time, e.g. higher-dimensional general relativity \cite{andriot2017signatures} and KK theory \cite{alesci2005can} etc. In this work, we will discuss all the possible polarization modes of linear gravitational waves in arbitrary dimensional space-time. Though there are still three types of polarization modes (tensor, vector and scalar) as the 4-dimensional space-time case, there are new polarization modes in tensor and vector modes, the number of which depend on the dimension of space-time.\par
Experimentally, current ground based detectors are not sensitive to non-tensoral modes, and have found no conclusive evidence for the existence or absence of these modes. However, the future space borne gravitational wave detectors and pulsar timing array detectors will have higher non-tensoral mode detection sensitivity. And many recent works \cite{Zhang:2021fha,Omiya:2020fvw,Xie:2022wkx} have analyzed this feasibility. This is one of the motivations for our discussion of observational effects in the second half of this paper.\par
In this work, we mainly focus on the classification of gravitational waves according to their polarizations in higher-dimensional space-time, and discuss the observation effects of different polarization modes of gravitational waves in the presence of extra dimensions. In section \ref{section:2}, we calculate all components of the Riemann tensor that could affect test particles in inertial reference by geodesic deviation equations. Each component could contribute to one or more of the possible polarization modes of gravitational waves. Both null and nonnull waves are discussed. Though nonnull wave case is more complicated compared to null wave case, the maximum number of possible independent polarization modes of gravitational wave is the same in both cases. The difference between the null wave and nonnull wave cases can appear when we do the Lorentz transformation. In section \ref{section:3} we classify null and nonnull gravitational waves in higher-dimensional space-time according to the Lorentz transform property of each polarization mode. In section \ref{section:4}, we mainly discuss the observation effect of the gravitational waves with the existence of extra dimensions. We calculate all possible polarization modes observed in 3-dimensional subspace for higher-dimensional gravitational waves and discuss two unique phenomena in the presence of extra dimensions, and show that the existence of these phenomena can serve as evidence for the existence of extra dimensions. 
\section{Polarizations of gravitational waves in $D$-dimensional space-time}\label{section:2}
In this section, we calculate all polarizations of gravitational waves in $D$-dimensional space-time, where we always set $D\geq 4$ in this work. It is worth noting that we only treat metric theories of gravity, in other words, gravity is only determined by the Riemann curvature tensor and a freely moving or falling particle always move along a geodesic. Apart from this, we do not make any restriction on gravity theories.

\subsection{Polarizations of null gravitational waves}
Before treating the polarizations of gravitational waves, let us briefly review how to detect gravitational waves, in other words how gravitational waves affect matters. We cannot detect gravitational waves by a single particle like what we usually do in the case of electromagnetic waves for the tensor nature of gravity. Thus, we must measure gravitational waves nonlocally, which means we need at least a pair of separated test particles. In order to describe this physical process, geodesic deviation equations are needed
\begin{equation}
\label{E1}
	\nabla_{\xi}\nabla_{\xi}N^{a}=R^{a}_{bcd}u^{b}u^{c}N^{d},
\end{equation}
where $\xi=\xi^{\alpha}\partial_{\alpha}$ is the velocity along the timelike geodesic $\gamma(\tau)$, and $N^{d}$ are the components of the separation vector between two infinitesimally separated geodesics. \par
For linear gravitational waves far away from sources, we can further simplify Eq. (\ref{E1}). The background space-time is Minkowski space-time. Thus, we can use partial derivative instead of covariant derivative in the radiation zone. Besides that we consider an inertial reference, for which each test particle is stationary in this frame. Thus, the geodesic deviation equations become
\begin{equation}
\label{E2}
	\frac{d^{2}N^{u}}{dt^{2}}=R^{u}_{\ tvt}N^{v},
\end{equation}
where here we use the representation in the Cartesian coordinate and the indices $u$ and $v$ represent $D-1$ spacial coordinates. \par
With the decomposition of the Riemann tensor  into traceless part and trace part
\begin{equation}
\label{E101}
\begin{split}
	R_{abcd}=&C_{abcd}+\frac{2}{D-2}(g_{a[c}R_{d]b}-g_{b[c}R_{d]a})\\&-\frac{2}{(D-1)(D-2)}Rg_{a[c}g_{d]b},
	\end{split}
\end{equation}
we can treat the Weyl tensor and the Ricci tensor separately. Then we replace the Cartesian frame $(e_{t}, e_{z}, m_{i})$ to the GHP one $(l, n, m_{i})$, and we always set the $(l, n)$ plane in the GHP formalism corresponds to $(e_{t}, e_{z})$ plane in the cartesian frame in this work. In addition, gravitational waves are propagating along $z$ direction in our assumption.  $l$ and $n$ respectively represent in going and out going waves.\par
 All the components of the Weyl tensor and the Ricci tensor can be classified by their boost weight in Table \ref{table1} and Table \ref{table2}. 
\begin{table}
\centering	
	\begin{tabular}{|c|c|c|c|}
		\hline Boost weight&Component&Notation&Spin\\
		\hline
		\hline 2 & $C_{lilj}$ & $\Omega_{ij}$&2\\
		\hline 1 & $C_{lijk}$&$\Psi_{ijk}$&3\\
		&$C_{lnli}$&$\Psi_{i}$&1\\
		\hline 0&$C_{ijkl}$&$\Phi_{ijkl}$&4\\
		&$C_{linj}$&$\Phi_{ij}$&2\\
		&$C_{lnij}$&$2\Phi_{ij}^{A}$&2\\
		&$C_{lnln}$&$\Phi$&0\\
		\hline -1& $C_{1ijk}$&$\Psi'_{ijk}$&3\\
		&$C_{nlni}$&$\Psi'_{i}$&1\\
		\hline -2&$C_{ninj}$&$\Omega'_{ij}$&2\\
		\hline
	\end{tabular}
	\caption{The classification of the Weyl tensor by its boost weight \cite{durkee2010generalization}.}
	\label{table1}
\end{table}
Where the indices $l, n, i$ represent $l, n, m_{i}$ in the GHP formalism. \par
\begin{table}
\centering	
	\begin{tabular}{|c|c|c|c|}
		\hline Boost weight&Component&Notation&Spin\\
		\hline
		\hline 2&$R_{ll}$&$\omega$&0\\
		\hline 1&$R_{li}$&$\psi_{i}$&1\\
		\hline 0&$R_{ij}$&$\phi_{ij}$&2\\
		 &$R_{ln}$&$\phi$&0\\
		 \hline -1&$R_{ni}$&$\psi'_{i}$&1\\
		 \hline -2&$R_{nn}$&$\omega'$&0\\
		 \hline
	\end{tabular}
	\caption{The classification of the Ricci tensor by its boost weight \cite{durkee2010generalization}.}
	\label{table2}
\end{table}
Now, when consider the geodesic deviation equations (\ref{E2}), we can filter out all components of the Weyl tensor and the Ricci tensor that could affect the test particles we choose. We list them in Table \ref{table3},
\begin{table*}
\centering	
	\begin{tabular}{|c|c|c|c|}
	\hline Boost weight&Component&Spin&Independent components\\
	\hline
	\hline 2&$\Omega_{ij}$&2&$\frac{1}{2}D(D-3)$\\
	&$\omega$&0&1\\
	\hline 1&$\Psi_{i}$&1&D-2\\
	&$\psi_{i}$&1&D-2\\
	\hline 0&$\Phi_{ij}$&2&$\frac{1}{2}D(D-3)$\\
	 &$\phi_{ij}$&2&$\frac{1}{2}(D-1)(D-2)$\\
	&$\Phi$&0&1\\
	&$\phi$&0&1\\
	\hline -1&$\Psi'_{i}$&1&D-2\\
	&$\psi'_{i}$&1&D-2\\
	\hline -2&$\omega'$&0&1\\
	&$\Omega'_{ij}$&2&$\frac{1}{2}D(D-3)$\\
	\hline		
	\end{tabular}
	\caption{All components that could affect the test particles.}
	\label{table3}
\end{table*}
\begin{table}
\centering
\begin{tabular}{|c|c|}
	\hline Cart.& GHP\\
	\hline
	\hline $R_{titj}$ & $R_{lilj},R_{linj},R_{nilj},R_{ninj}$\\
	\hline $R_{titz}$ & $R_{liln},R_{niln}$\\
	\hline $R_{tztj}$ & $R_{lnlj},R_{lnnj}$\\
	\hline $R_{tztz}$ & $R_{lnln}$\\
	\hline
\end{tabular}
\caption{The relationship between components of the Riemann curvature tensor in the Cartesian coordinate and the GHP coordinate.}
\label{table6}	
\end{table}
where the numbers of independent components in the last column of Table \ref{table3} are calculated from their symmetry, e.g.  $\Omega_{ij}$ satisfies $\Omega_{ij}=\Omega_{ji}$, $\Omega_{ii}=0$. For convenience in relating the polarization modes to the components of the Riemann curvature tensor when discussing the polarization modes later, here we decompose the GHP components in Table \ref{table6} into  traceless part and trace part
\begin{equation}
	\begin{array}{lr}
	\begin{split}
		R_{linj}=&C_{linj}+\frac{1}{D-2}(\delta_{ij}R_{ln}+R_{ij})+\\&\frac{1}{(D-2)(D-1)}\delta_{ij}R,
		\end{split}\\
		R_{lilj}=C_{lilj}+\frac{1}{D-2}\delta_{ij}R_{ll},\\
		R_{liln}=C_{liln}+\frac{1}{D-2}R_{li},\\
		R_{lnln}=C_{lnln}+\frac{2}{D-2}R_{ln}-\frac{1}{(D-1)(D-2)}R.\\
	\end{array}
\end{equation}
where $C_{linj}\equiv \Phi_{ij}$, $C_{lilj}\equiv \Omega_{ij}$, $C_{liln}\equiv\Psi_{i}$, $C_{lnln}\equiv\Phi$, $R_{ln}\equiv\phi$, $R_{ij}\equiv\phi_{ij}$, $R\equiv2\phi+\phi^{i}_{i}$, $R_{ll}\equiv\omega$ and $R_{li}\equiv\psi_{i}$ in the GHP symbols. 
And by swapping $l$ and $n$ with each other we can obtain all the decompositions of GHP components in Table \ref{table6}.\par
Finally we have enough background knowledge and it is time to discuss the polarizations of null gravitational waves. \par
For null plane gravitational waves, wave functions are only determined by their retarded time $t-z$. Thus, the curvature tensor satisfies
\begin{equation}
	R_{abcd,p}=0,
\end{equation}
where $a, b, c$ represent $l, n, m_{i}$ and $p, q$ represent $l, m_{i}$. Then with the Bianchi identity
\begin{equation}
	R_{ab[pq,n]}=\frac{1}{3}R_{abpq,n}=0,
\end{equation}
we have
\begin{equation}
	R_{abpq}=0.
\end{equation}
\par
Therefore, all nonzero components have the form $R_{anbn}$ \cite{eardley1973gravitational}. Comparing with Table \ref{table3} and Table \ref{table6}, we have the following nonzero components for null gravitational waves
\begin{equation}
\label{E31}
	\Omega'_{ij}, R_{ninl}, R_{lnln}, \omega',
\end{equation}
when $D=4$, these notations correspond to the common notations
\begin{equation}
	\Omega'_{ij}\longrightarrow\Psi_{4}, R_{ninl}\longrightarrow\Psi_{3}, R_{lnln}\longrightarrow\Psi_{2}, \omega'\longrightarrow\Phi_{22}.
\end{equation}
\par
For the convenience of analysis, we write $R_{tutv}$ in the matrix form (i.e. the polarization matrix) in the Cartesian coordinate $(e_{t}, e_{z}, m_{i})$
\begin{equation}
\label{E39}
R_{tutv}=
\left(
\begin{matrix}
	
	F_{ij}(\Omega'_{ij},\omega')&V_{i}(R_{ninl})\\
	& \\
	V_{j}(R_{njnl}) &S(R_{lnln})\\
\end{matrix}
\right)	,
\end{equation}
where the $z$ axis (corresponding to the last row and last column of the above polarization matrix) parallel to the propagation direction, $F_{ij}(\Omega'_{ij},\omega')=\frac{1}{4}(\Omega'_{ij}+\frac{1}{D-2}\delta_{ij}\omega')$ is the transverse part of the above polarization matrix with the traceless part (the tensor modes) $\Omega'_{ij}$ and the trace part (breathing scalar mode) $\omega'$, $V_{i}(R_{ninl})=\frac{1}{4}R_{ninl}$ are the vector modes and $S(R_{lnln})=\frac{1}{4}R_{lnln}$ is the longitudinal scalar mode. Such a classification is a generalization of the 4-dimensional case.\par
Next, according to the analysis method in Ref. \cite{szekeres1965gravitational,podolsky2012interpreting}, we can further classify these components according to their specific effect on the test particles in Table \ref{table10},
\begin{table}
	\centering
	\begin{tabular}{|c|c|}
	\hline $\Omega'_{ij}$&transverse effect\\
	\hline $R_{ninl}$ &longitudinal effect\\
	\hline $R_{lnln}, \omega'$ &Newton-Coulomb deformations\\
	\hline	
	\end{tabular}
	\caption{The classification of the Riemann curvature components by their effect on test particles \cite{szekeres1965gravitational,podolsky2012interpreting}.}
	\label{table10}
\end{table}
where $\Omega'_{ij}$ are all the nonzero components in Einstein's theory and they have $\frac{1}{2}D(D-3)$  independent components. For a general metric gravity theory, the number of independent components listed in (\ref{E31}) is at most $\frac{1}{2}D(D-1)$.

\subsection{Polarizations of nonnull gravitational waves}
In this part, we generalize the above conclusion to nonnull waves, which are the gravitational waves in massive gravity theories. In Ref. \cite{hyun2019exact}, the authors calculated the nonzero components of the curvature tensor in 4-dimensional space-time. Here we generalize the result to $D$-dimensional space-time.\par
We still set the propagation direction to the $z$ axis. So any wave vector is a linear combination of $l$ and $n$. Unlike the null wave case, the directions of a nonnull wave vector will change with the Lorentz boost transformation. So by using the similar method used in the null wave case, we have $R_{abpq}=0$ and $p,q$ represent $i$. Thus, all the components of the Riemann tensor which could affect test particles in Table \ref{table3} and Table \ref{table6} could be nonzero. In contrast to it, there are some components must be zero in null cases according to the analysis in last subsection. In fact, these nonzero components corresponds to some polarizations. We list them here
\begin{equation}
\begin{array}{lr}
	R_{lilj},R_{linj},R_{ninj},\\R_{liln},R_{niln},R_{lnlj},R_{lnnj},\\R_{lnln}.
	\end{array}
\end{equation} 
Then by doing the same analysis as the previous subsection, we can write $R_{tutv}$ in the matrix form in the Cartesian coordinate $(e_{t}, e_{z}, m_{i})$

\begin{equation}
\label{E40}
R_{tutv}=
\left(
\begin{matrix}
	F'_{ij}(R_{lilj},R_{linj},R_{ninj}) &V'_{i}(R_{liln},R_{niln})\\
	& \\
	V'_{j}(R_{lnlj},R_{lnnj}) &S'(R_{lnln})\\
\end{matrix}
\right)	,
\end{equation}
where $F'_{ij}(R_{lilj},R_{linj},R_{ninj})=\frac{1}{4}(R_{linj}+R_{lilj}+R_{ninj})$, $V'_{i}(R_{liln},R_{niln})=\frac{1}{4}(R_{liln}+R_{niln})$ and $S'(R_{lnln})=\frac{1}{4}R_{lnln}$.\par 
Just the same as the null wave case, we can still divided the above polarization matrix into the following  parts:\\
the terms contributing to tensor modes (the traceless part of $F'_{ij}(R_{lilj},R_{linj},R_{ninj})$ on $i$ and $j$ indices) are
\begin{equation}
	R_{linj}^{T}\equiv R_{linj}-\frac{1}{D-2}g_{ij}g^{kr}R_{lknr},\Omega_{ij},\Omega'_{ij};
\end{equation}
the terms contributing to vector modes are
\begin{equation}
	R_{liln},R_{niln};
\end{equation}
the terms contributing to breathing mode (the trace
 part of $F'_{ij}(R_{lilj},R_{linj},R_{ninj})$ on $i$ and $j$ indices) are
\begin{equation}
	R_{linj}^{S}\equiv g^{ij}R_{linj},\omega\equiv R_{ll},\omega'\equiv R_{nn};
\end{equation}
the term contributing to longitudinal mode is
\begin{equation}
	R_{lnln}.
\end{equation}
\par
For the transverse traceless tensor modes, there are $3\times \big(\frac{1}{2}D(D-3)\big)$ components. For the vector modes, there are $2\times (D-2)$ components. And for the scalar modes, there are $4$ components. Note that they are not independent.\par
In conclusion, according to the geodesic deviation equations, only electric part ($R_{tutv}$) of the Riemann curvature tensor can affect the test particles for linear waves. We can always divide the Riemann curvature tensor into tensor modes, vector modes and scalar modes three parts and we always have $\big(\frac{1}{2}D(D-3)+(D-2)+2\big)$ independent freedoms no matter the gravitational waves are null or not. But there are more components that can affect the polarizations if the waves are nonnull. They can bring about complexity in their classification and observation. 

\section{Classification of gravitational waves}\label{section:3}
In the previous section, we calculated all components of the Riemann tensor for linear plane gravitational waves which could affect the test particles in Minkowski space-time. In this section, we mainly discuss a classification of these components.
The classification of these components by Lorentz transformation was first mentioned in Ref. \cite{eardley1973gravitational}, the authors proposed $E(2)$ classification to null plane waves in 4-dimensional space-time. 
 The advantage of this classification is that any element belongs to one and only one of the classes. Physically, which means that for each gravitational waves with any combination of polarization modes (tensor, vector ,breathing and longitudinal) belongs to one and only one of the classes (for nonnull case, we do not make this assumption since the situation is complicated, we will discuss it in the last part of this section). In this section we generalize it to null and nonnull linear plane waves in $D$-dimensional space-time. 
\subsection{Classification of null gravitational waves}
Now, we study the classification of null plane gravitational waves in $D$-dimensional space-time. This requires us to study the transformation behavior of the polarization modes of the null plane gravitational waves in $D$-dimensional space-time under Lorentz transformation. The 4-dimensional case has been studied in Ref. \cite{eardley1973gravitational}. Ref. \cite{eardley1973gravitational} pointed out that we can just consider the subgroup of $SO(3,1)$ that meets the following two assumptions without losing generality: (1) each observer sees that wave is traveling in his $+z$ direction; (2) the frequency of a monochromatic gravitational wave remains unchanged before and after the transformation. This is because after any Lorentz transformation: For (1), the gravitational wave can always propagate along the positive direction of the $z$ axis through a pure space rotation. A pure space rotation does not change the polarization modes of the gravitational waves. For (2), the frequency of a gravitational wave can always be changed by a boost along the propagation direction. A pure boost along the propagation direction does not change the polarization modes of gravitational waves either. Therefore, we can always discuss the classification of the Riemann curvature tensor under the subgroup of the entire Lorentz group.  It is worth noting that these two assumptions are unnecessary. Even though dropping them would make the subgroup larger, this would not affect our classification of the components of the Riemann curvature tensor, we make this assumption only to simplify the calculation. So for the nonnull case we will discuss in next subsection, we only use the first assumption that the direction of the wave vector for each inertial observer is parallel to the $z$ axis. According to this constraint, the classification applicable to all inertial observers can be obtained.
\par
In 4-dimensional space-time, the NP formalism has been used to calculate the classification of null gravitational waves in Ref. \cite{eardley1973gravitational}. In the case of null gravitational waves, the most general group of Lorentz transformation that satisfies the two assumption in the previous paragraph is the subgroup that keeps the null vector $l^{a}$ unchanged. This transformation can be written in spinor form as
\begin{equation}
   \begin{array}{lr}
   	o^{A} \mapsto e^{i\frac{\varphi}{2}}o^{A},\\
   	\iota^{A} \mapsto e^{-i\frac{\varphi}{2}}(\iota^{A}+\bar{c}o^{A}),
   \end{array}
\end{equation}
where $\varphi$ is the rotation angle in the 2-plane $m\times \bar{m}$ and $c\in \mathbb{C}$. Rewrite the transformations in the NP formalism, we have
\begin{equation}
	\begin{array}{lr}
		l^{a} \mapsto l^{a},\\
		n^{a} \mapsto n^{a}+\bar{c}m^{a}+c\bar{m}^{a}+c\bar{c}l^{a},\\
		m^{a} \mapsto e^{i\varphi}(m^{a}+cl^{a}),\\
		\bar{m}^{a} \mapsto e^{-i\varphi}(\bar{m}^{a}+\bar{c}l^{a}),\\
	\end{array}
\end{equation}
where $l$ is the special constant null vector we choose. Now we are going to deal with the higher-dimensional case. Though the spinor formalism is very useful and simple in 4-dimensional space-time, but it loses its advantage in higher-dimensional space-time. So we use the GHP formalism instead.

 In the $D$-dimensional GHP formalism, the most general Lorentz transformations  that keep $l$ fixed are
\begin{equation}
	\begin{array}{lr}
	l\mapsto l,\\
	n\mapsto n+c_{i}m_{i}-\frac{1}{2}c^{2}l,\\
	m_{i}\mapsto X_{ij}(m_{j}-c_{j}l),\\	
	\end{array}
\end{equation}
where $c_{i}$ are $D-2$ arbitrary constants, $c^{2}=c_{i}c_{i}$ and $X_{ij}$ are the rotation matrix in $m_{2}\times ...\times m_{D-1}$ subspace. Note that $X_{ij}$ are also directly related to spin, and by this we distinguish between tensor-vector and scalar modes in the last section.\par
So by doing the above transformations on the Riemann tensor we have
\begin{equation}
	\begin{array}{lr}
		R_{lnln} \mapsto R_{lnln},\\
		R_{ninl} \mapsto X_{ij}(R_{njnl}-c_{j}R_{lnln}),\\
		\Omega'_{ij} \mapsto X_{ik}X_{jr}(\Omega'_{kr}-2c_{(k}R_{nr)nl}+c_{k}c_{r}R_{lnln}),\\
		\omega' \mapsto \omega'+2c_{i}\psi'_{i}-
		c^{2}\phi.\\
	\end{array}
\end{equation}
\par
Therefore all the polarizations of null plane gravitational waves in $D$-dimensional space-time can be classified to the following six classes: \par
Class$O_{0}$. $\Omega'_{ij}= R_{ninl}= R_{lnln}= \omega'=0$, there is no wave for all inertial observers.\par
Class$O_{1}$. $\Omega'_{ij}= R_{ninl}= R_{lnln}=0, \omega'\neq 0$, all inertial observers agree with the present of $\omega'$.\par
Class$N_{2}$. $\Omega'_{ij}\neq0, R_{ninl}= R_{lnln}= \omega'=0$, all inertial observers agree with the present of $\Omega'_{ij}$.\par
Class$N_{3}$. $R_{ninl}= R_{lnln}=0, \Omega'_{ij}\neq0\neq\omega'$, all inertial observers agree with the present of $\Omega'_{ij}$ and $\omega'$.\par
Class$III_{4}$. $R_{lnln}=0\neq R_{ninl}$, all inertial observers agree with the present of $R_{ninl}$ and the absence of $R_{lnln}$, all the other components are observers dependent.\par
Class$D_{5}$. $R_{lnln}\neq0$, all inertial observers agree with the present of $R_{lnln}$ and all the other components are observers dependent. \par
\begin{table}[h!]
\centering	
	\begin{tabular}{|c||c|c|c|c|}
	\hline Class&Tensor&Breathing&Vector&Longitudinal\\
	\hline
	\hline Class$O_{0}$ &&&&\\
	\hline Class$O_{1}$ &&\checkmark &&\\
	\hline Class$N_{2}$ &\checkmark &&&\\
		\hline Class$N_{3}$ &\checkmark &\checkmark &&\\
		\hline Class$III_{4}$ &?&?&\checkmark &\\
		\hline Class$D_{5}$ &?&?&?&\checkmark \\
\hline
	\end{tabular}
	\caption{Polarization modes of each classification of null gravitational waves. Here``?" means the present of the corresponding mode depends on observers and ``$\checkmark$" means the present of the corresponding mode is agreed by all observers. Blank space means the absence of the corresponding mode is agreed by all observers.}
\label{table8}
\end{table}
Here we summarize the classification results in the Table \ref{table8}.\par
For any metric gravity theory, linear null plane gravitational waves must belong to one of the above six classes. The higher-dimensional classification and the 4-dimensional one are quite similar in form.

\subsection{Classification of nonnull gravitational waves}
The classification of nonnull gravitational waves is more complicated than that of null ones. Since the velocity of a nonnull gravitational wave is less than the speed of light, its wave vector is observer dependent and can be given by $\frac{l+sn}{\sqrt{2s}}$, where $s\in \mathbb{R}$.
Here we can set the propagation direction for each inertial observer parallel to the $z$ axis. Thus, the only constraint condition is that the wave vector for any inertial observer is located in $(l, n)$ plane.
 Next, we discuss the classification in 4-dimensional space-time first and then extend it to $D$-dimensional space-time.\par
In 4-dimensional case, we still use the NP formalism. In order to obtain the most general Lorentz transformation, we set
\begin{equation}
   \begin{array}{lr}
   	o^{A} \mapsto e^{i\frac{\varphi}{2}}(ao^{A}+b\iota^{A}),\\
   	\iota^{A} \mapsto e^{-i\frac{\varphi}{2}}(co^{A}+d\iota^{A}),
   \end{array}
\end{equation}
where $a,b,c,d$ in these subsection are arbitrary complex numbers which satisfy $ad-bc=1$ and $\bar{a}\bar{d}-\bar{b}\bar{c}=1$ due to Eq. (\ref{E32}).
In the case of null waves, we impose two assumptions on the Lorentz transformation. We have pointed out that those two assumptions are not necessary. They are just for us to consider fewer transformations without losing generality, so as to simplify the calculation. However, the two assumptions of the null wave case do not apply to the nonnull case. Instead of these two assumptions, we use the following assumption: the propagation direction for any observer parallels to the $z$ axis. Thus, we have $a\bar{b}+sc\bar{d}=0$ and $ \bar{a}b+s\bar{c}d=0$.
  These are all the constraints we can obtain. Then by rewriting these transformations in the NP formalism, we have
  \begin{equation}
  \label{E42}
	\begin{array}{lr}
		l^{a} \mapsto a\bar{a}l^{a}+b\bar{b}n^{a}+a\bar{b}m^{a}+\bar{a}b\bar{m}^{a},\\
		n^{a} \mapsto c\bar{c}l^{a}+d\bar{d}n^{a}+c\bar{d}m^{a}+\bar{c}d\bar{m}^{a},\\
		m^{a} \mapsto a\bar{c}l^{a}+b\bar{d}n^{a}+a\bar{d}m^{a}+\bar{c}b\bar{m}^{a},\\
		\bar{m}^{a} \mapsto c\bar{a}l^{a}+d\bar{b}n^{a}+b\bar{c}m^{a}+\bar{a}d\bar{m}^{a},\\
	\end{array}
\end{equation}
where the parameters satisfy the following constraints 
\begin{equation}
\label{E44}
	\begin{array}{lr}
		ad-bc=1,\\
		\bar{a}\bar{d}-\bar{b}\bar{c}=1,\\
		a\bar{b}+sc\bar{d}=0,\\
		\bar{a}b+s\bar{c}d=0.\\
	\end{array}
\end{equation}
\par
 And now we are going to generalize it to $D$-dimensional case. Just the same as the null case, the $D$-dimensional transformation is just a simple generalization of 4-dimen-sional one. Since we have obtained all the possible nonzero components in last section,
by doing the most general Lorentz transformation on these components we can obtain
\begin{equation}
\label{E20}
	\begin{array}{lr}
	\begin{split}
	\Omega'_{ij}\mapsto &X_{ik}X_{jl}\times 
f_{1kl}(\Omega'_{kl},\Omega_{kl},R_{liln},R_{niln},R_{lnln},R_{linj}^{S}),
	\end{split}\\
	\begin{split}
	\Omega_{ij}\mapsto &X_{ik}X_{jl}\times 
f_{2kl}(\Omega'_{kl},\Omega_{kl},R_{liln},R_{niln},R_{lnln},R_{linj}^{S}),
	\end{split}\\
	\begin{split}
	R_{linj}^{T}\mapsto &X_{ik}X_{jl}\times 
f_{3kl}(R_{linj}^{T},R_{liln},R_{niln},R_{lnln},\omega', \omega),
	\end{split}\\
	\begin{split}
	R_{liln}\mapsto &X_{ik}\times 
f_{4k}(\Omega'_{kl},\Omega_{kl},R_{linj}^{T},R_{liln},R_{niln},\\&R_{lnln},\omega', \omega,R_{linj}^{S}),
	\end{split}\\
	\begin{split}
	R_{niln}\mapsto &X_{ik}\times 
f_{5k}(\Omega'_{kl},\Omega_{kl},R_{linj}^{T},R_{liln},R_{niln},\\&R_{lnln},\omega', \omega,R_{linj}^{S}),
	\end{split}\\
	\begin{split}
	R_{lnln}\mapsto & 
f_{6}(\Omega'_{kl},\Omega_{kl},R_{linj}^{T},R_{liln},R_{niln},R_{lnln},\\&\omega', \omega,R_{linj}^{S}),
	\end{split}\\
	\begin{split}
	\omega'\mapsto  
f_{7}(R_{linj}^{T},R_{liln},R_{niln},R_{lnln},\omega', \omega,R_{linj}^{S}),
	\end{split}\\
	\begin{split}
	\omega \mapsto  
f_{8}(R_{linj}^{T},R_{liln},R_{niln},R_{lnln},\omega', \omega,R_{linj}^{S}),
	\end{split}\\
	\begin{split}
	R_{linj}^{S}\mapsto & 
f_{9}(R_{linj}^{T},R_{liln},R_{niln},R_{lnln},\omega', \omega,R_{linj}^{S}),
	\end{split}\\
	\end{array}
\end{equation}
where $f_{n}$ represent the linear functions of respective variables. Here we ignore lengthy calculations, just write out the functional relationship qualitatively, since the subsequent analysis does not require specific quantitative calculations of Eq. (\ref{E20}). The detailed equations are placed in the Appendix \ref{appendix:1}. Now, we have enough information to discuss the classification: \par
Class$O_{0}$. All components are equal to zero for an inertial observer and all the other inertial observers agree with it.\par
Class$O_{r}$. One of $\omega'$, $ [\omega]$ and $( R_{linj}^{S})$ is not equal to zero and all the other components are equal to zero, all the other inertial observers agree on the present of this one. The present of components in $R_{linj}^{T},R_{liln},R_{niln},R_{lnln},$ $ \omega,R_{linj}^{S}$, $[R_{linj}^{T},R_{liln},R_{niln},R_{lnln}, \omega',R_{linj}^{S}]$ and $(\Omega'_{kl},\Omega_{kl},$ $R_{liln},R_{niln},R_{lnln},\omega', \omega,R_{linj}^{S})$ are observer dependent. \par
Class$N_{s}$. One of $\Omega'_{ij}$ and $( \Omega_{ij})$ is not equal to zero and all the other components are equal to zero, all the other inertial observers agree on the present of this one. The present of any components in $\Omega_{kl},R_{liln},R_{niln},R_{lnln}$ and $(\Omega'_{kl},R_{liln},R_{niln},R_{lnln})$ are observer dependent.\par
Class$N_{t}$. $R_{linj}^{T}$ is not equal to zero and all the other components are equal to zero, all the other inertial observers agree on the present of this one. The present of any components in $R_{liln},R_{niln},R_{lnln}, \omega',\omega,R_{linj}^{S}$ are observer dependent.\par
Class$III_{yi}$. One of $R_{liln}, R_{niln}$ is not equal to zero and all the other components are equal to zero, all the other inertial observers agree on the present of this one. The present of any other component is observer dependent.\par
Class$D_{y}$. $R_{lnln}$ is not equal to zero and all the other components are equal to zero, all the other inertial observers agree on the present of this one. The present of any other component is observer dependent.\par
Note that any combination of Class$O_{r}$, Class$N_{s}$, Class$N_{t}$, Class$III_{yi}$ and Class$D_{y}$ is a new classification. \par
\begin{table}[h!]
\centering	
	\begin{tabular}{|c||c|c|c|c|}
	\hline Class&Tensor&Breathing&Vector&Longitudinal\\
	\hline
	\hline Class$O_{0}$ &&&&\\
	\hline Class$O_{r}$ &?&\checkmark &?&?\\
	\hline Class$N_{s}$ &\checkmark &&?&?\\
		\hline Class$N_{t}$ &\checkmark &? &?&?\\
		\hline Class$III_{yi}$ &?&?&\checkmark &?\\
		\hline Class$D_{y}$ &?&?&?&\checkmark \\
\hline
	\end{tabular}
	\caption{Polarization modes of each classification of nonnull gravitational waves.}
\label{table9}
\end{table}

For the nonnull case, our treatment is slightly different from the null case. We only list what modes could be observed for any observer in the present of each mode. Thus, the combination of two or more classes can represent a new class. Therefore, one of the best ways is to calculate the specific form of Eq. (\ref{E20}) for a specific theory, and then we can analyze the amplitude of each polarization mode for any observer.\par


\section{The possibility of observation}\label{section:4}

In this section, we study how different types of gravitational waves in higher dimensions affect the observations in our 3-dimensional space. We start from the simplest case of analyzing the observational effects of each higher-dimensional polarization mode in a 3-dimensional subspace. Then we use the classification in section \ref{section:3} to find the polarization modes for each type of gravitational wave. Next, we give two phenomena that are common in extra-dimensional theories, but rare in 4-dimensional modified gravity. Therefore, in the detection of gravitational waves with higher precision in the future, these phenomena can be used as the criterion for the existence of extra dimensions.\par
In the final part, we broaden the applicability of our results. We begin with the most general warped extra-dimensional metric and apply the method in this paper \cite{andriot2017signatures} to show that the two phenomena we present are still valid in any Ricci-flat extra-dimensional scenario.

 \subsection{Observation effect for each type mode in higher-dimensional space-time}
\par
For any higher-dimensional gravitational wave, let us first discuss which polarization modes could be observed in the 3-dimensional subspace. Extra-dimensional theories can be classified from the curvature perspective into flat extra-dimensional theories (KK theory \cite{kaluza1921unity}, domain wall model \cite{rubakov1983we}, DGP model \cite{dvali20004d} and large extra dimensional brane model \cite{arkani1998hierarchy} etc.) and warped extra-dimensional theories (warped extra dimensional brane models \cite{randall1999large,randall1999alternative} etc.), and from the topological perspective into infinite extra dimensions (domain wall model \cite{rubakov1983we}, DGP model \cite{dvali20004d} and Randall-Sundrum II model \cite{randall1999alternative} etc.) and compact extra dimensions (KK theory \cite{kaluza1921unity}, large extra dimensional brane model \cite{arkani1998hierarchy} and Randall-Sundrum I model \cite{randall1999large}). In this subsection, we only discuss the case of flat extra dimensions. In the last subsection, we will generalize some of the results to the case of arbitrary Ricci-flat warped extra dimensions.\par
 Note that in the case of infinitely large flat extra dimensions, the space-time is higher-dimensional Minkowski at large scales, so we can analyze the properties of gravitational waves in higher-dimensional Minkowski background, for example, DGP theory \cite{Khlopunov:2022ubp,Khlopunov:2022jaw}. This method can also be generalized to the case of compact extra dimensions,
the compactness of extra dimensions does not affect our conclusions, since compaction is equivalent to adding periodic conditions to extra dimensions. Compaction only discretizes the extra dimensions' components of wave vectors and does not affect the polarization modes. \par
\begin{figure}
	\centering
	\includegraphics[width=4cm]{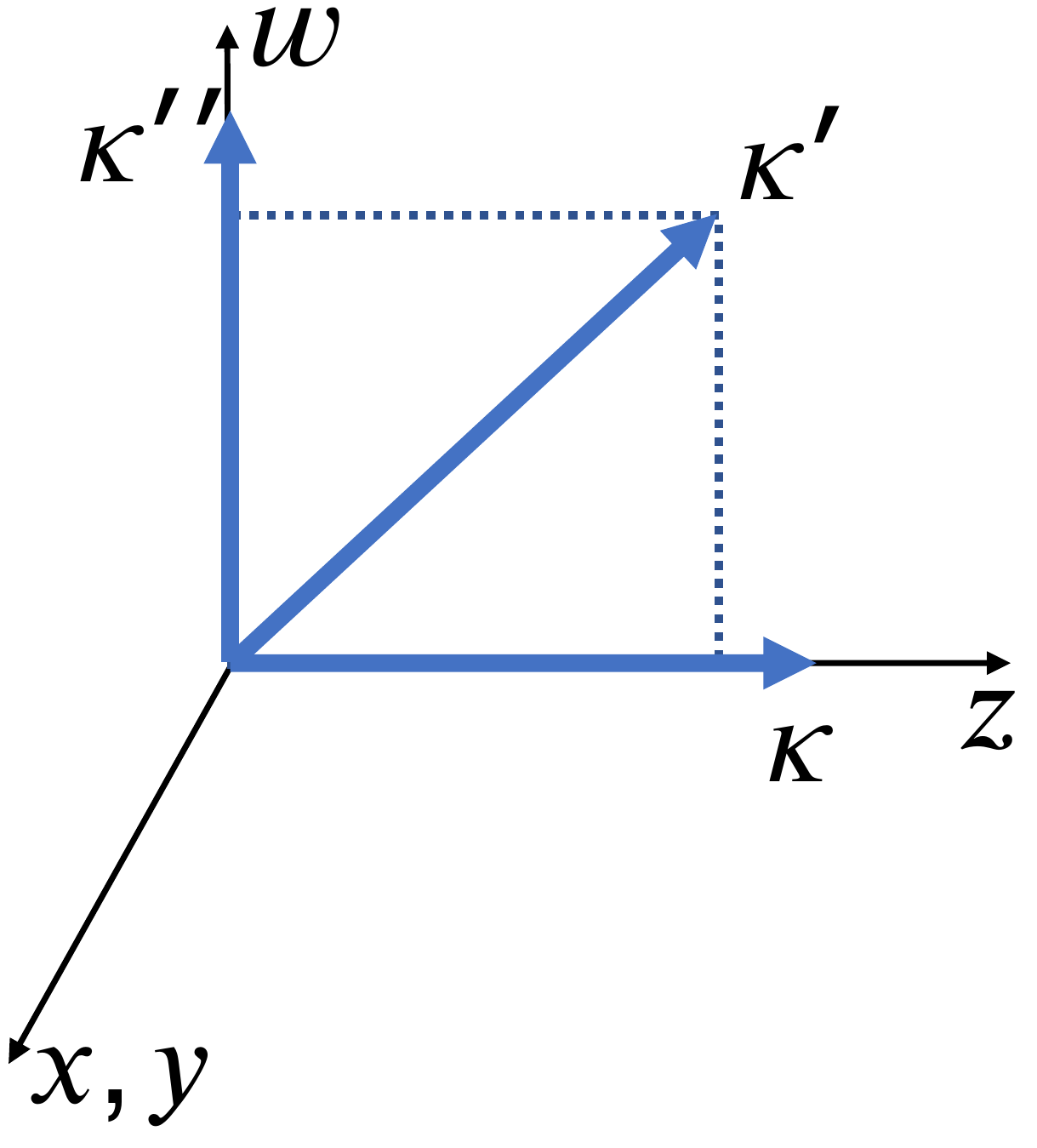}
	\caption{Three typical wave vectors of higher-dimensional space-time gravitational waves. $\kappa$, $\kappa'$ and $\kappa''$ represent the direction of wave vectors in \textit{case 1}, \textit{case 2} and \textit{case 3} respectively. $w$ is one of the extra dimensions.}
	\label{p2}
\end{figure}

 To analyze which polarization modes could be observed in the 3-dimensional subspace we need to project the polarization tensor of gravitational waves in higher-dimensional space onto the 3-dimensional subspace which we live in. Without loss of generality, we take $(e_{x},e_{y},e_{z})$ as the basis of the 3-dimensional subspace. Then for all plane gravitational waves propagating in higher-dimensional space-time, they can be divided into three cases based on their wave vectors' direction. In first case the waves only propagate in 3-dimensional subspace. In second case the waves propagate in both 3-dimensional subspace and extra dimensions, and in third case the waves only propagate in extra dimensions. For convenience, we use \textit{case 1}, \textit{case 2} and \textit{case 3} to represent the above three cases respectively, see FIG. \ref{p2}. Different cases correspond to different types of polarization matrices. For flat extra dimensions, the projection of the polarization matrix (see Eq. (\ref{E39}) and Eq. (\ref{E40}) correspond to null case and nonnull case respectively) onto the 3-dimensional subspace is to intercept the 3-dimensional part of the whole matrix to obtain a third-order polarization matrix. We mainly discuss the first two cases here, because in \textit{case 3}, we cannot define the propagation direction of gravitational waves in the 3-dimensional subspace. It is worth noting that here we still take the projection of any wave vector in the 3-dimensional subspace as the $z$-axis without loss of generality. When we check all possible cases, we can obtain all the polarization modes which could be observed in 3-dimensional subspace for higher-dimensional gravitational waves, see Table \ref{table11}. \par
\begin{figure}
	\centering
	\includegraphics[width=5cm]{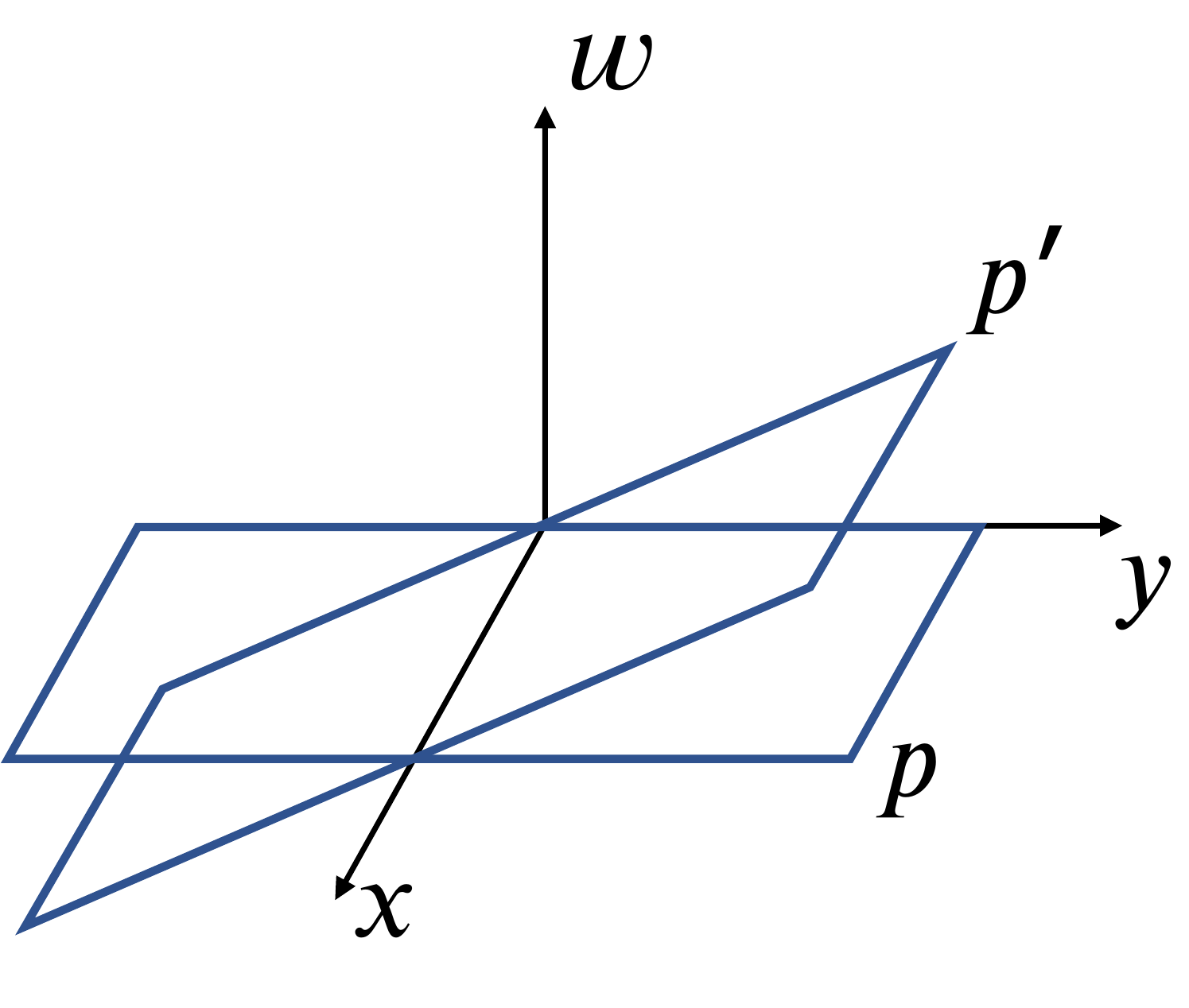}
	\caption{The possible location of a 2-dimensional polarization plane. $p$ and $p'$ represent two cases.}
	\label{p3}
\end{figure}

\begin{figure}
	\centering
	\includegraphics[width=3cm]{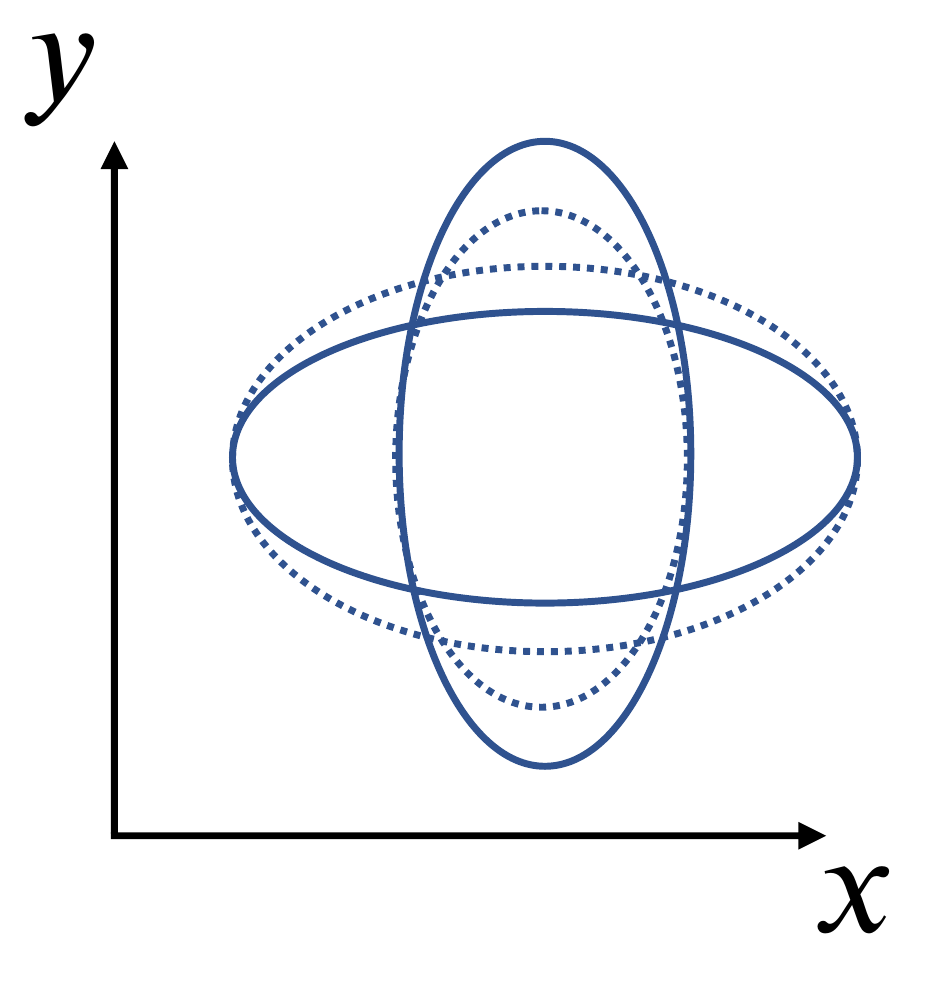}
	\caption{Schematic diagram of the response of ($+$) mode in 3-dimensional subspace of the two cases in FIG. \ref{p3}. The solid and dashed lines represent the polarization modes that we can observe in the 3-dimensional subspace ($e_{x},e_{y},e_{z}$) for the two 2-planes $p$ and $p'$, respectively.}
	\label{p4}
\end{figure}

 As an example, we take one of the tensor modes in $D$-dimensional space-time that only affects the test particles in a 2-dimensional polarization plane. This is similar to that the ($+$) mode and ($\times$) mode in 4-dimensional space-time affect the test particles in the ($x,y$) plane. There are more than one plane for the higher-dimensional space-time case. Our observation on this tensor mode in 3-dimensional subspace depends on the position of this 2-dimensional polarization plane. One of the possibilities is that it is a subspace of 3-dimensional subspace. We use $p$ to represent this kind of 2-dimensional polarization plane in FIG. \ref{p3}. This is exactly the same as the situation where there is no extra dimensions, e.g. $(+)$ mode in the 2-plane $(e_{x}, e_{y})$. Another possibility is just like $p'$ in FIG. \ref{p3}, i.e. there are nonzero metric perturbations in both 3-dimensional subspace and extra dimensions. For example, we make a rotation on the polarization plane $p$ to obtain a new polarization plane $p'$, see FIG. \ref{p3}. This process can be represented by a transformation of the polarization matrix:
 \begin{equation}
 \label{E34}
 \begin{split}
 	&p:\left(
 	\begin{matrix}
 		h_{+} &0&0&0\\
 		0& -h_{+} &0&0\\
 		0&0&0&0\\
 		0&0&0&0\\
 	\end{matrix}
 	\right)
 	\mapsto p':S^{-1}\left(
 	\begin{matrix}
 		h_{+} &0&0&0\\
 		0& -h_{+} &0&0\\
 		0&0&0&0\\
 		0&0&0&0\\
 	\end{matrix}
 	\right)S\\&
 	=\left(
 	\begin{matrix}
 		h_{+} &0&0&0\\
 		0& -h_{+}\cos^{2}\theta &0&-h_{+}\sin \theta \cos\theta \\
 		0&0&0&0\\
 		0&-h_{+}\sin \theta \cos\theta &0&-h_{+}\sin^{2}\theta \\
 	\end{matrix}
 	\right),\\&
 	S=\left(
 	\begin{matrix}
 		1 &0&0&0\\
 		0& \cos \theta &0&\sin \theta \\
 		0&0&1&0\\
 		0&-\sin \theta &0&\cos \theta \\
 	\end{matrix}
 	\right),
 	\end{split}
 \end{equation}
 where we use only one extra dimension $w$  for convenient. The spacial basis here is ($e_{x},e_{y},e_{z},e_{w}$).
  In this case, the amplitudes of the polarization in two perpendicular directions in the 2-plane $(e_{x}, e_{y})$ are different when $-h_{yy}=h_{+}\cos^{2}\theta$ is not equal to $h_{xx}=h_{+}$. And this leads to an extra breathing mode for the observers in 3-dimensional subspace. The amplitude, phase and frequency of this breathing mode completely depend on the corresponding tensor mode. The relationship between them only depend on the position of the 2-dimensional polarization plane $p'$, see FIG. \ref{p4}. \par

\begin{table*}[htbp]
\centering	
	\begin{tabular}{|c|cccccc|cccccc|} 
	\hline \multirow{2}{*}{Polarization}&\multicolumn{6}{c|}{\textit{Case 1}}&\multicolumn{6}{c|}{\textit{Case 2}} \\
	 &+&$\times$&b&x&y&l&+&$\times$&b&x&y&l\\
	\hline Tensor &\checkmark &\checkmark &\checkmark &&&&\checkmark &\checkmark &\checkmark &\checkmark &\checkmark &\checkmark\\
	\hline Breathing &  & &\checkmark  &&&&  &  &\checkmark  & & &\checkmark  \\
	\hline Longitudinal & && &&&\checkmark &  &  &  & & &\checkmark  \\
	\hline Vector & && &\checkmark &\checkmark &&  &  &  &\checkmark &\checkmark &  \\
	\hline
		\end{tabular}
	\caption{The observable polarizations in 3-dimensional subspace. Any ``$\checkmark$'' mode listed in the table could be observed in the 3-dimensional subspace. The first column lists polarization modes of higher-dimensional gravitational waves. The second and third columns list possible polarization modes observed in 3-dimensional space in \textit{case 1} and \textit{case 2}, respectively. ``$+$'' and ``$\times$'' represent two tensor modes, ``$b$'' represents breathing mode, ``$x$'' and ``$y$'' represent two vector modes, ``$l$'' represents longitudinal mode. }
	\label{table11}
\end{table*}
\par

\subsection{Observation Effect for Each Class of Gravitational Waves}
Now, we can apply the above result to the classification of gravitational waves obtained in the previous sections, and to determine how many polarizations we can observe for each class of gravitational wave in the present of extra dimensions. In the first part of this section, we divide gravitational waves into three cases (\textit{case 1}, \textit{case 2} and \textit{case 3}) according to whether the wave vectors have extra dimensions' components. For each class of gravitational wave we only consider \textit{case 1} and \textit{case 2}. Table \ref{table4} and Table \ref{table5} show the polarization modes of null and nonnull gravitational waves, respectively.

\begin{table*}[htbp]
\centering	
	\begin{tabular}{|c|cccccc|cccccc|}
	\hline \multirow{2}{*}{Class}&\multicolumn{6}{c|}{\textit{Case 1}}&\multicolumn{6}{c|}{\textit{Case 2}} \\
	 &+&$\times$&b&x&y&l&+&$\times$&b&x&y&l\\	
	 \hline Class$O_{0}$ &&&&&&&&&&&&\\
	\hline Class$O_{1}$ &&&\checkmark &&&& & &\checkmark &&&? \\
	\hline Class$N_{2}$ &\checkmark &\checkmark &? &&&&\checkmark &\checkmark &? &? &? &?\\
		\hline Class$N_{3}$ &\checkmark &\checkmark &\checkmark &&&&\checkmark &\checkmark &\checkmark &? &? &?\\
		\hline Class$III_{4}$ &?&?&?&\checkmark &\checkmark &&?&?&?&\checkmark &\checkmark &?\\
		\hline Class$D_{5}$ &?&?&?&? &? &\checkmark &?&?&?&? &? &\checkmark\\
\hline
	\end{tabular}
	\caption{Polarization modes of each classification of null gravitational wave.}
	\label{table4}
\end{table*}
The first column in Table \ref{table4} lists classes of polarization modes of null gravitational waves. The second and third columns respectively represent all the polarizations could be observed for \textit{case 1} and \textit{case 2}. Here in Class$O_{1}$, there is only one degree of freedom. For Class$N_{2}$, whether the breathing mode and longitudinal mode in \textit{case 2} are independent depends on dimension of space-time. When the dimension of the whole space-time is higher than 5, they are independent. The property that the different modes are not independent we discussed above is independent of specific theory, that is, it is true for all theories of tensor gravity. And for all the other classes, a theory can always be found that makes all modes independent.
\par

\begin{table*}[h!]
	\centering
	\begin{tabular}{|c|cccccc|cccccc|}
		\hline \multirow{2}{*}{Class}&\multicolumn{6}{c|}{\textit{Case 1}}&\multicolumn{6}{c|}{\textit{Case 2}} \\
	 &+&$\times$&b&x&y&l&+&$\times$&b&x&y&l\\
	 		\hline Class$O_{0}$ &&&&&&&&&&&&\\
		\hline Class$O_{r}$ &?&?&\checkmark &?&?&?&?&?&\checkmark &?&?&?\\
		\hline Class$N_{s}$ &\checkmark &\checkmark &? &?&?&?&\checkmark &\checkmark &?&? &? &?\\
		\hline Class$N_{t}$ &\checkmark &\checkmark &? &?&?&?&\checkmark &\checkmark &? &? &? &?\\
		\hline Class$III_{yi}$ &?&?&?&\checkmark &\checkmark &?&?&?&?&\checkmark &\checkmark &?\\
		\hline Class$D_{y}$ &?&?&?&?&?&\checkmark &?&?&?&?&?&\checkmark \\
		\hline
		
	\end{tabular}
	\caption{Polarization modes of each classification of nonnull gravitational wave.}
	\label{table5}
\end{table*}
Table \ref{table5} has the same structure as Table \ref{table4}, it summarizes all the polarizations of each class of nonnull wave. For Class$N_{s}$, whether the breathing mode and longitudinal mode in \textit{case 1} and \textit{case 2} are independent depends on dimension of the space-time. When the dimension of the whole space-time is higher than 5, they are independent.
\par
The above two tables include all possible cases in any metric gravity theory. Here we give a few examples.\par
The first example is general relativity in $D$-dimensional space-time. Note that there are still only tensor modes among the polarizations of gravitational waves in higher-dimensional space-time, which means it belongs to class$N_{2}$. But if we observe gravitational waves in a 3-dimensional subspace, some tensor modes can be recognized as vector modes or scalar modes. The number of all independent degree of freedoms is $\frac{1}{2}D(D-3)$. Thus, according to the forth line in Table \ref{table4}, if the propagation direction of the waves is still in the 3-dimensional subspace (\textit{case 1}), there might be one breathing mode in addition to the ($+$) and ($\times$) modes. When the gravitational waves are partly propagate into the extra dimensions (\textit{case 2}), all the polarization modes can be observed. It is worth noting that whether the two scalar modes (breathing mode and longitudinal mode) are independent depends on the dimension of the whole space-time. They are independent when the dimension $D$ is higher than 5.\par
The second example is Horndeski theory \cite{horndeski1974second}.
 In addition to the tensor modes, Horndeski theory also include massive or massless scalar modes. The massive scalar field excites a mix mode of the longitudinal and breathing modes, while the massless scalar field only has the breathing mode. Thus, gravitational waves in Horndeski theory belong to a mixed class. The null waves belong to Class$N_{3}$, and the nonnull waves belong to Class$O_{r}$ and Class$III_{y}$.\par

\subsection{Observation Effect on Detector}

In this subsection we calculate the response of gravitational waves on the detector with and without extra dimensions. This reveals the observational differences between four-dimensional gravity theory and higher-dimensional gravity theory, which we will summarize as two phenomena in the next subsection. We will point out that there are phenomena that are common and natural in extra dimensions but are difficult to explain in modified gravities in 4-dimensional space-time, so once we detect these phenomena there may be extra dimensions. For the convenience of calculation, we use the Michelson interferometer detector with two arms perpendicular to each other. We set the detector tensor as
\begin{equation}
	D_{ij}=e_{x}\otimes e_{x}-e_{y}\otimes e_{y},
\end{equation}
i.e. the two arms of the detector are parallel to the $x$-axis and the $y$-axis, respectively. And $D_{ij}$ contracting with the polarization tensor $E^{ij}$ is the response of the detector
\begin{equation}
	F=D_{ij}E^{ij}.
\end{equation}
\par

\begin{figure}[h!]
	\centering
	\subfigure[4-$D$ $(+)$ mode]{\includegraphics[width=2.5cm]{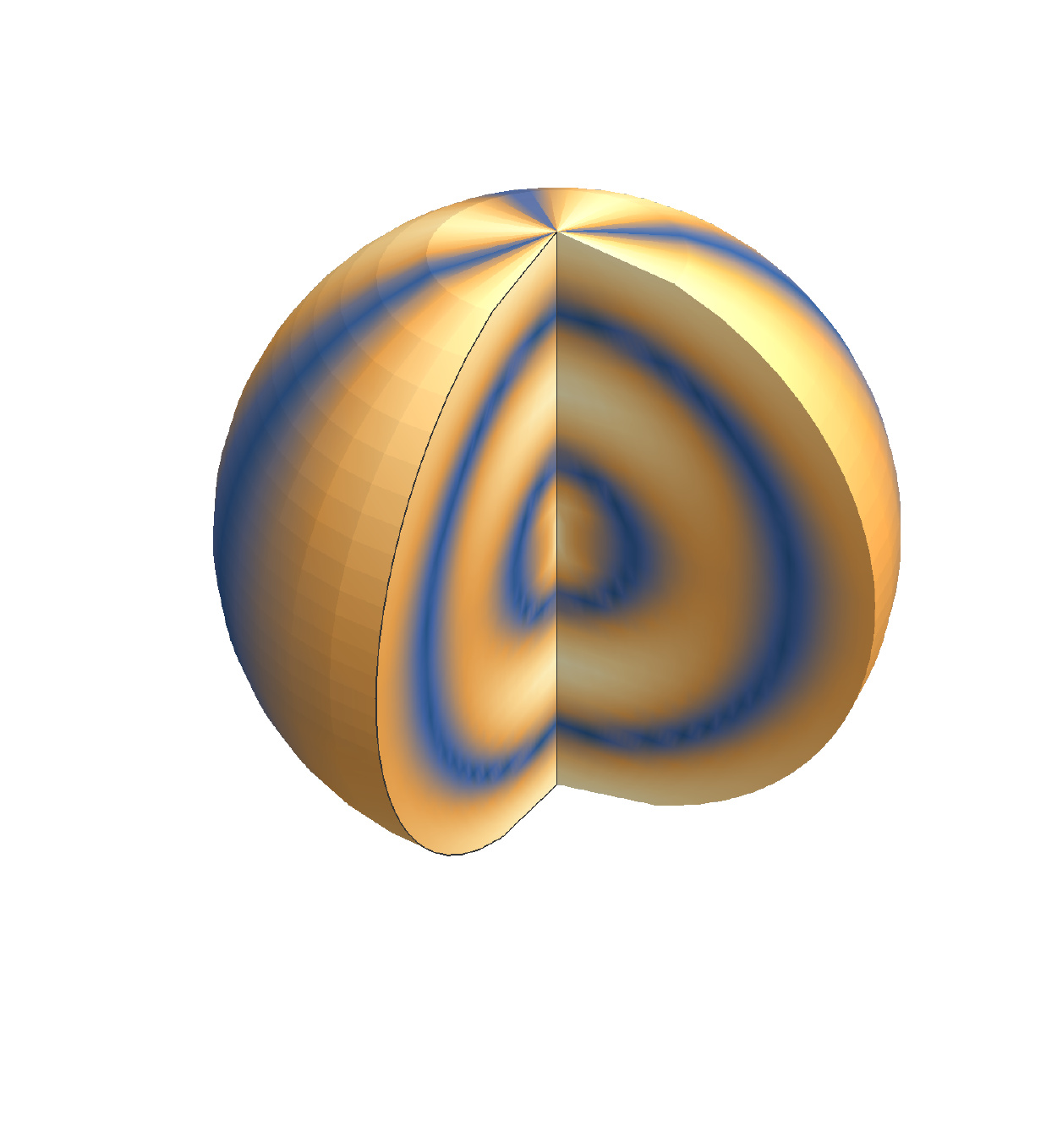}}
	\subfigure[(4+n)-$D$ $(+)$ mode]{\begin{minipage}[t]{0.12\textwidth}
		\includegraphics[width=2.5cm]{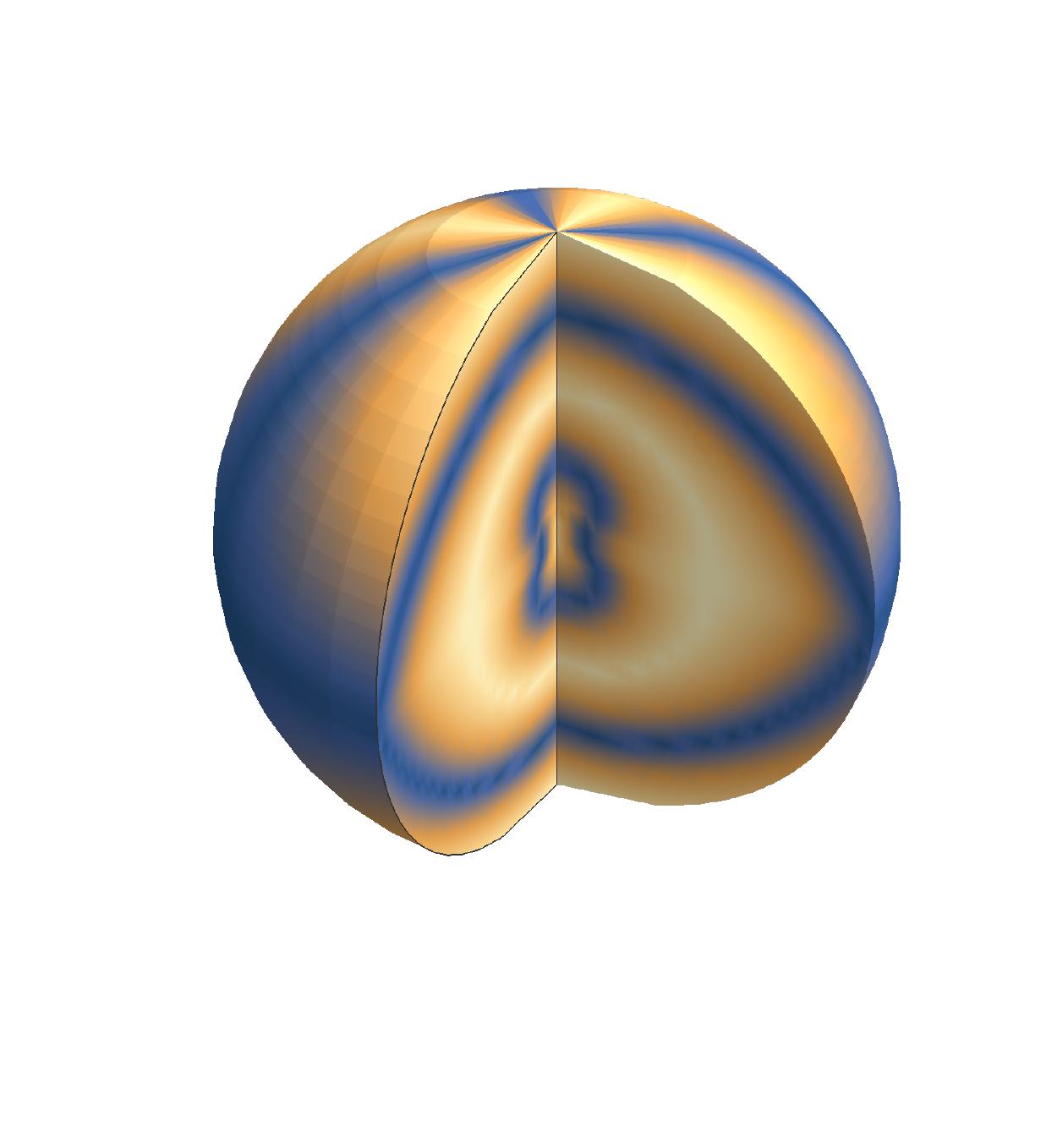}
	\end{minipage}
	\begin{minipage}[t]{0.12\textwidth}

		\includegraphics[width=2.5cm]{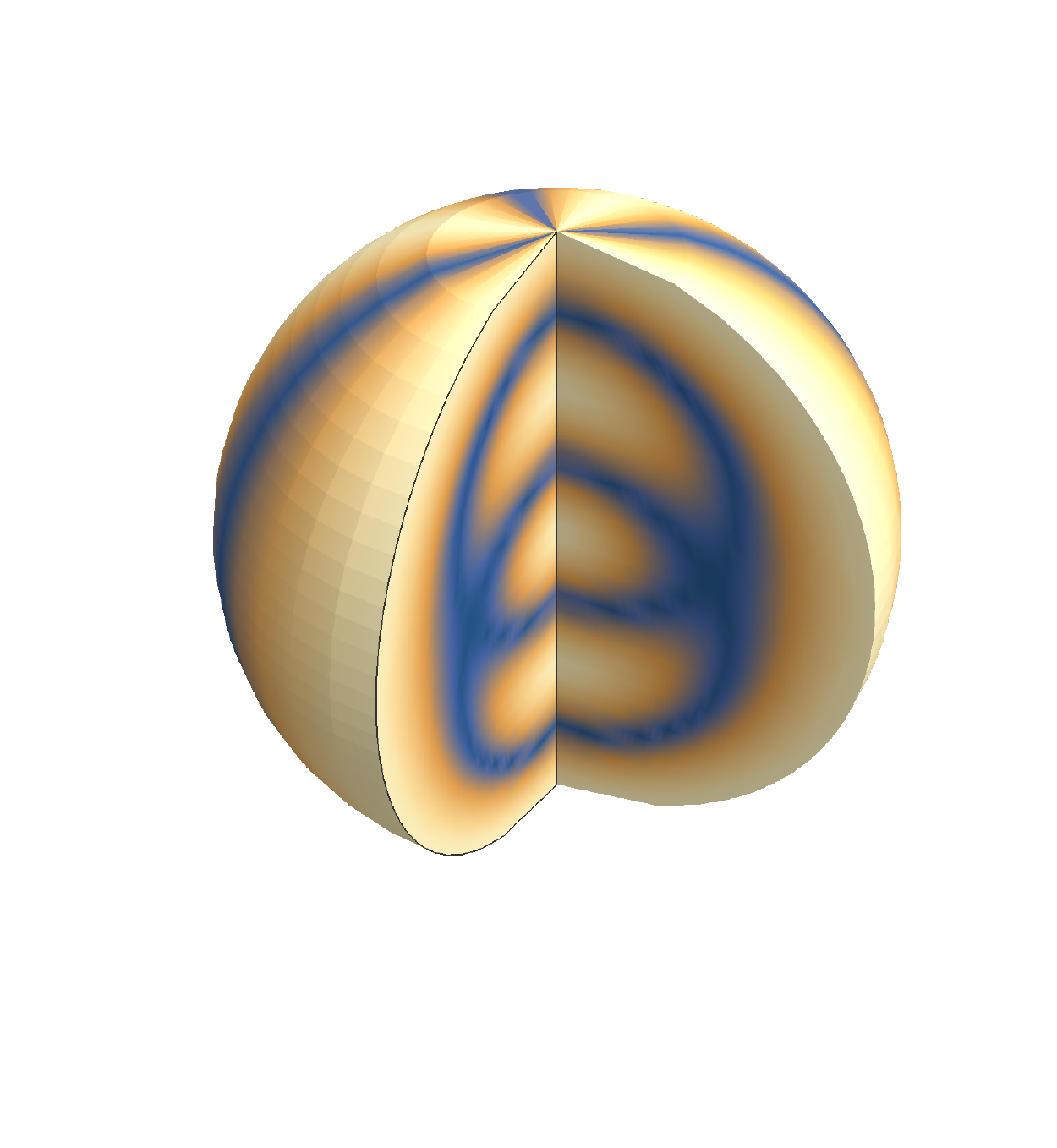}
	\end{minipage}}
	\\
	\subfigure[4-$D$ $(\times)$ mode]{\includegraphics[width=2.5cm]{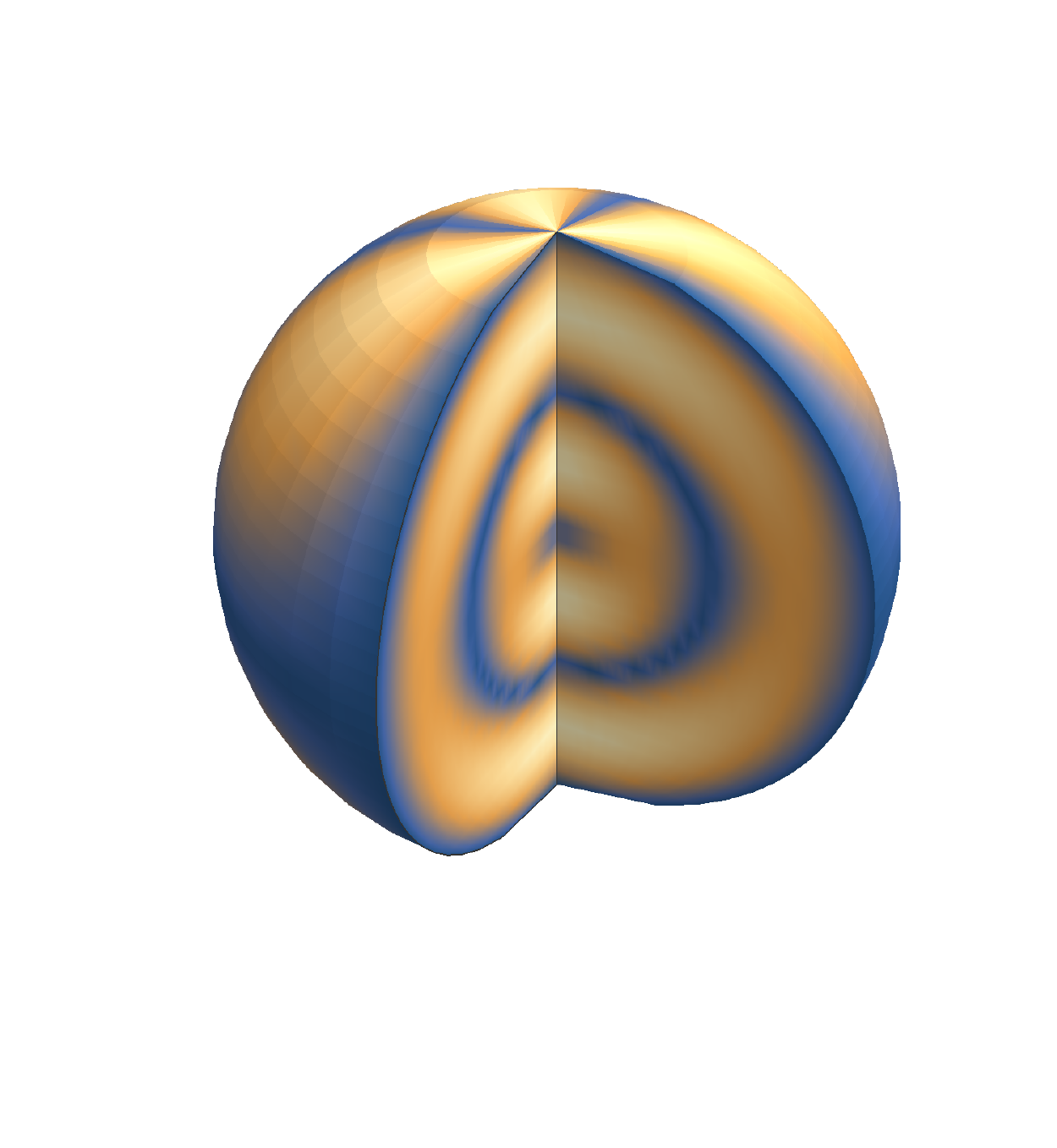}}
	\subfigure[(4+n)-$D$ $(\times)$ mode]{\begin{minipage}[t]{0.12\textwidth}
		\includegraphics[width=2.5cm]{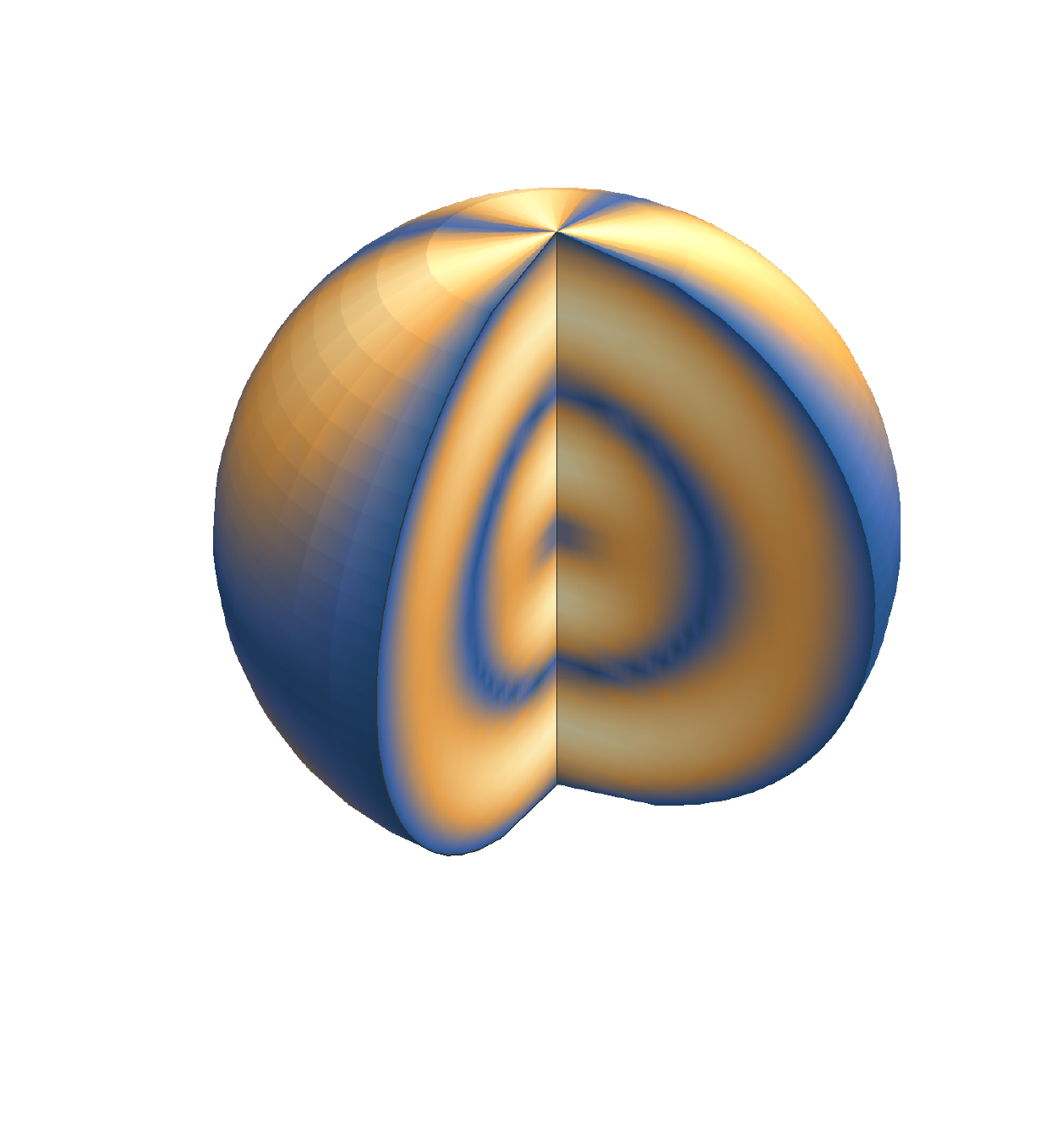}
	\end{minipage}
	\begin{minipage}[t]{0.12\textwidth}

		\includegraphics[width=2.5cm]{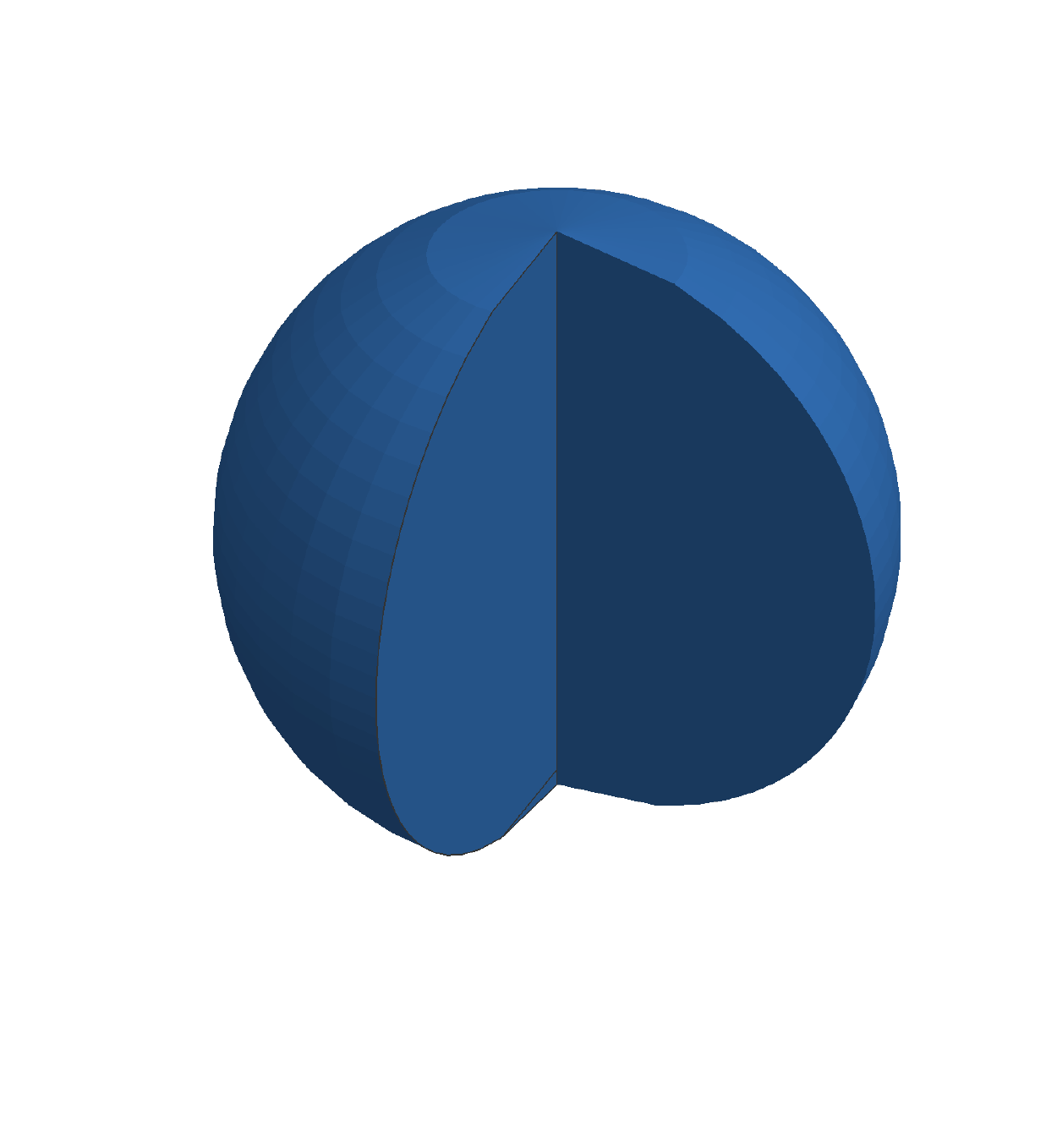}
	\end{minipage}}
	\\
	\subfigure[4-$D$ $(b)$ mode]{\includegraphics[width=2.5cm]{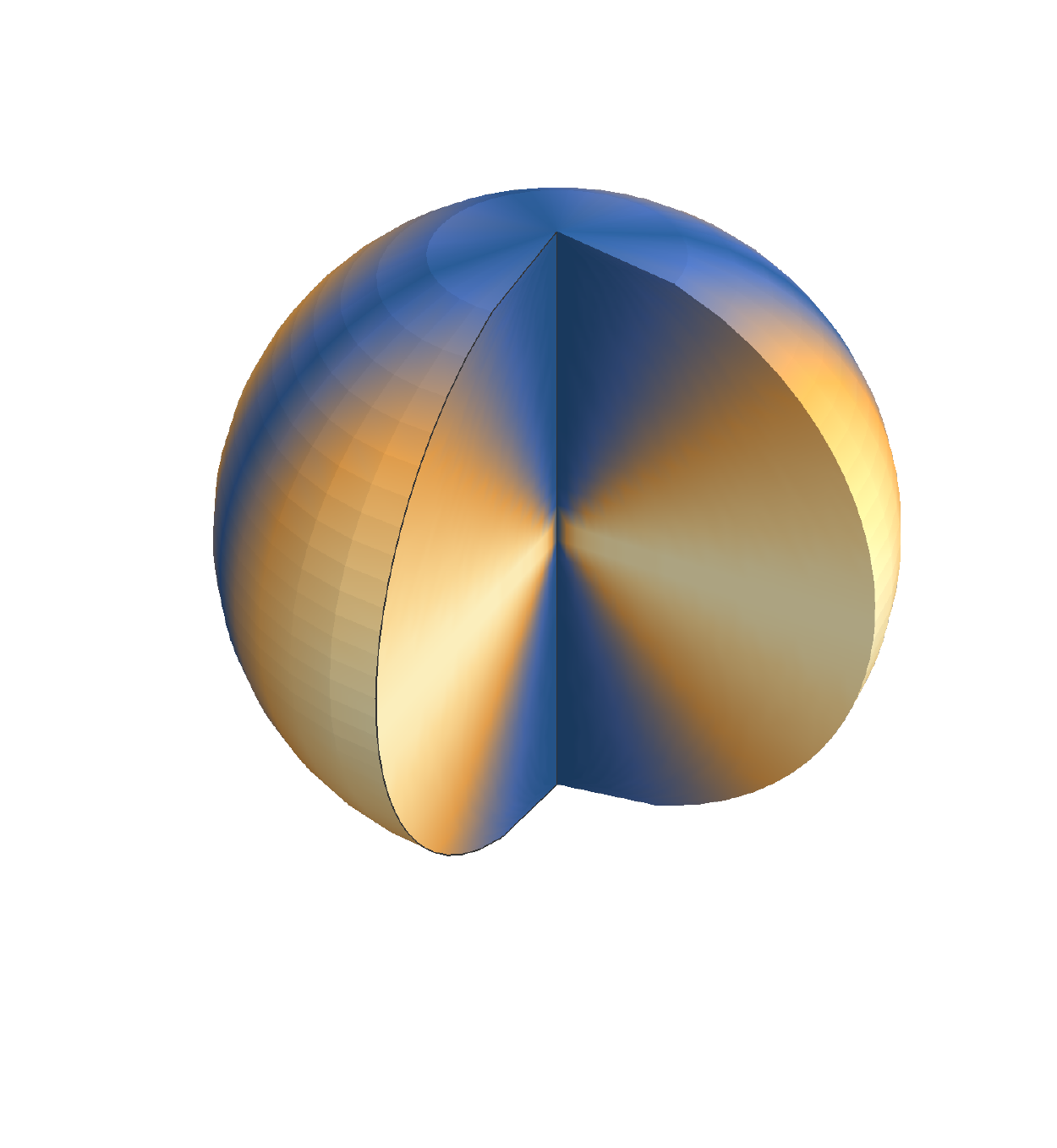}}
	\subfigure[(4+n)-$D$ $(b)$ mode]{\begin{minipage}[t]{0.12\textwidth}
		\includegraphics[width=2.5cm]{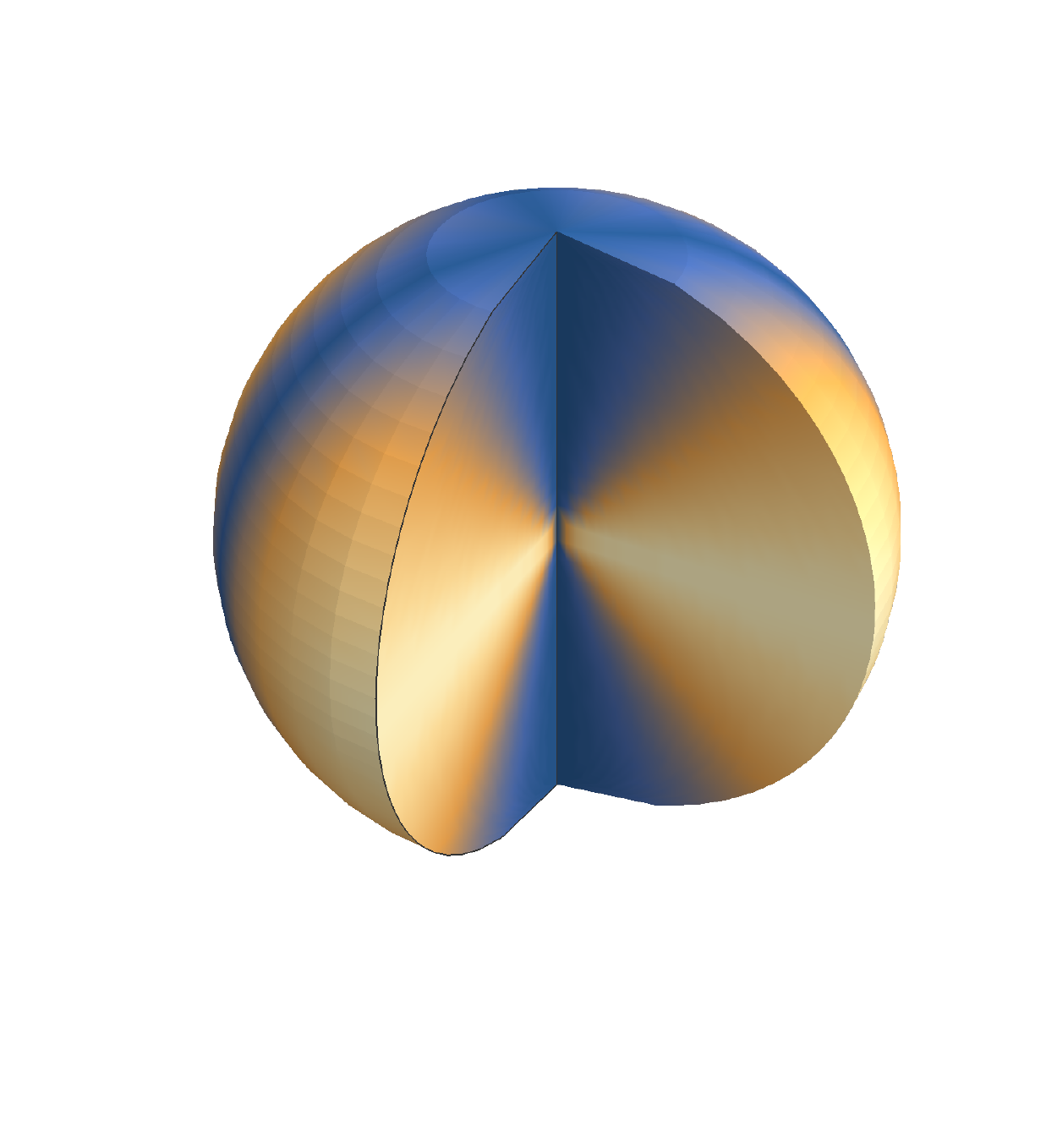}
	\end{minipage}
	\begin{minipage}[t]{0.12\textwidth}

		\includegraphics[width=2.5cm]{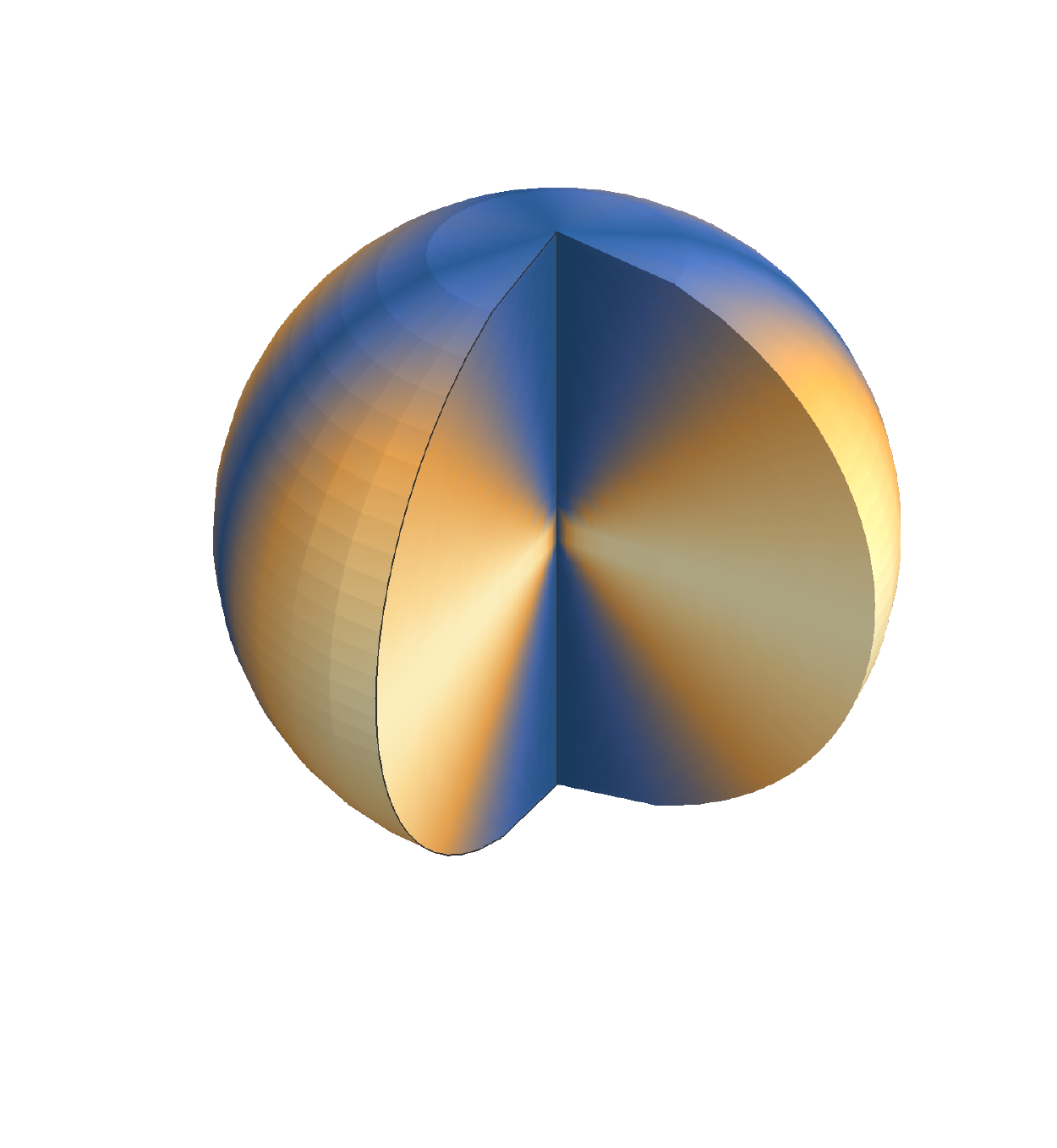}
	\end{minipage}}
	\\
	\subfigure[4-$D$ $(l)$ mode]{\includegraphics[width=2.5cm]{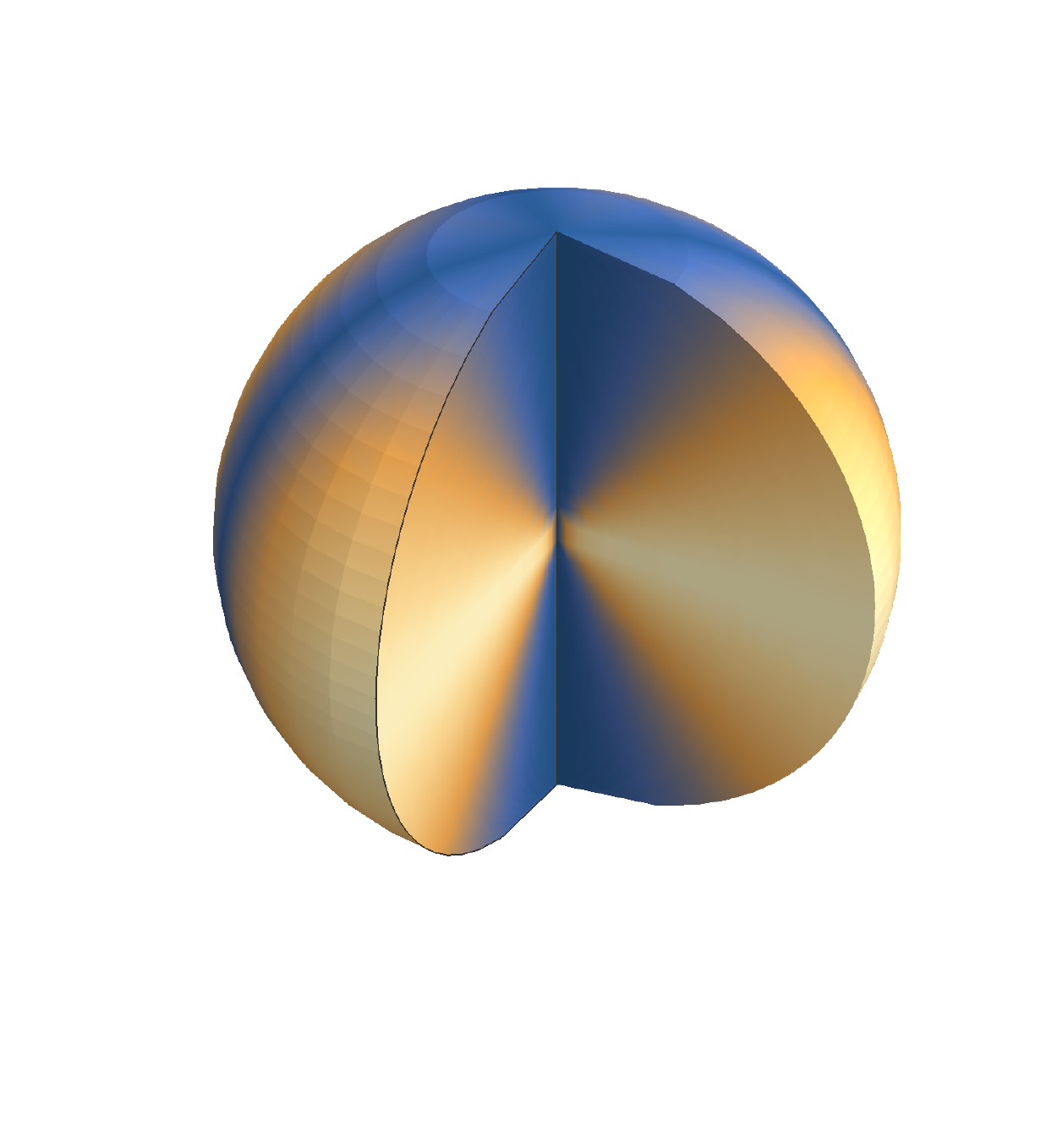}}
	\subfigure[(4+n)-$D$ $(l)$ mode]{\begin{minipage}[t]{0.12\textwidth}
		\includegraphics[width=2.5cm]{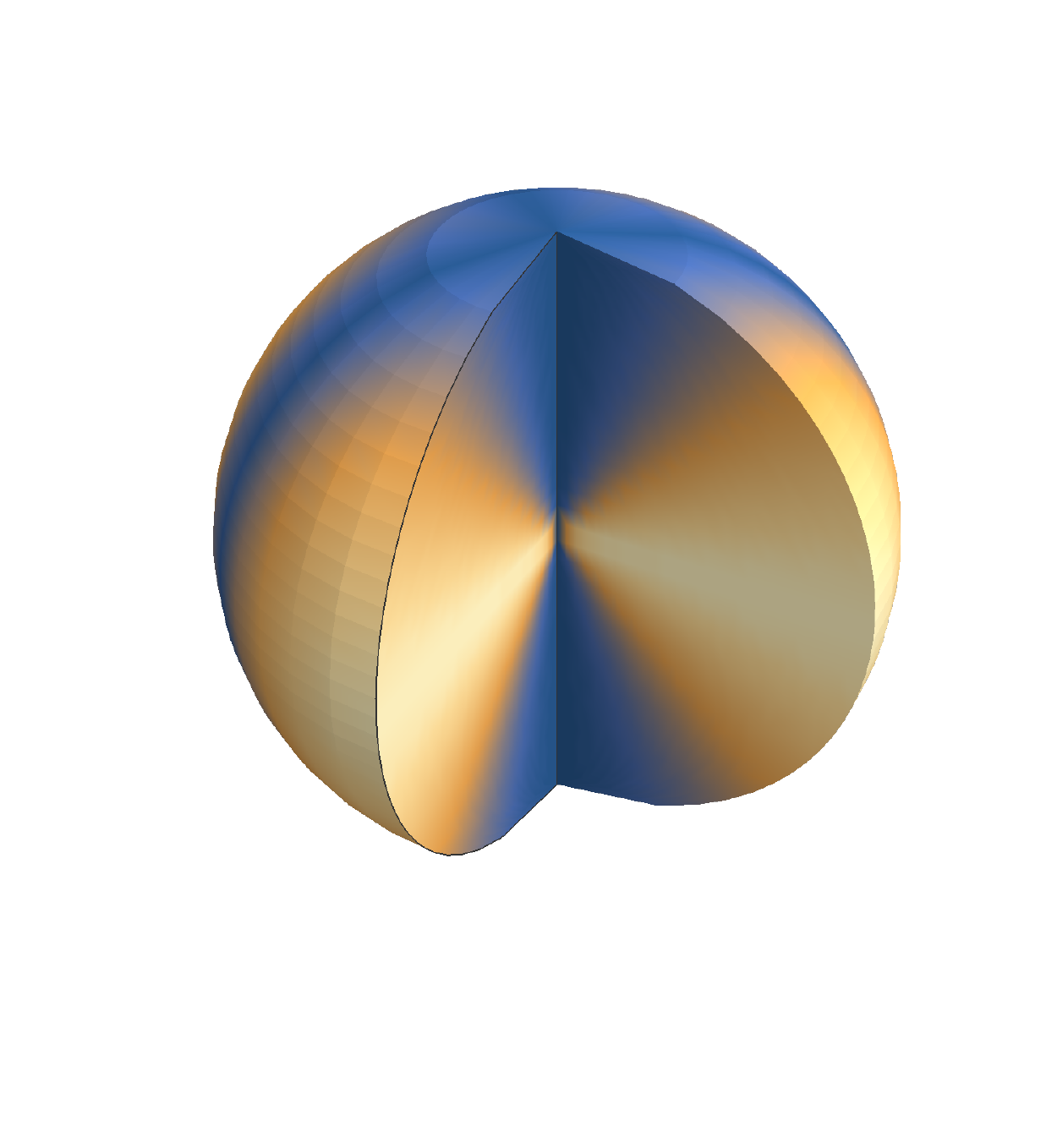}
	\end{minipage}
	\begin{minipage}[t]{0.12\textwidth}

		\includegraphics[width=2.5cm]{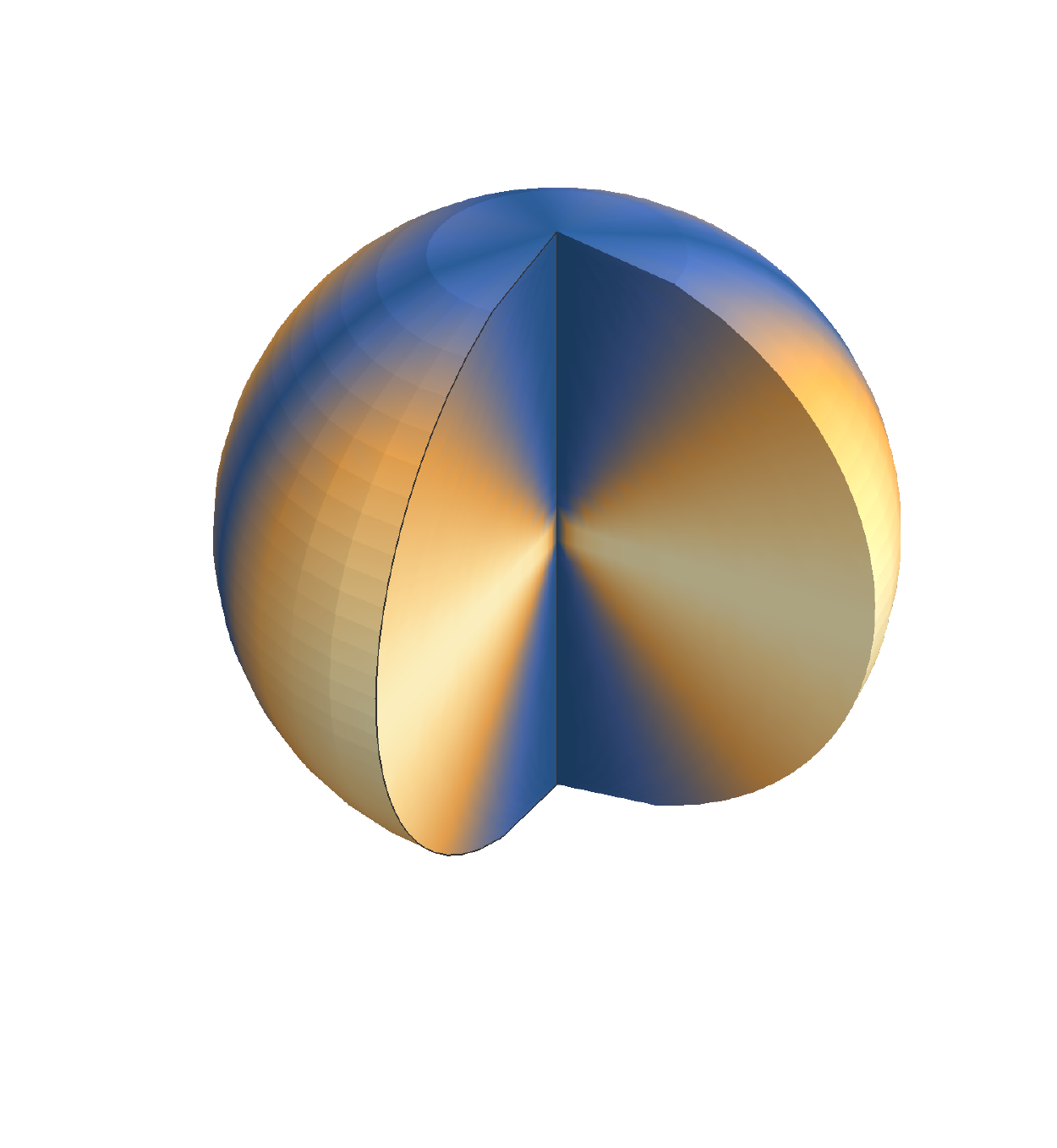}
	\end{minipage}}
	\\
	\subfigure[4-$D$ $(x)$ mode]{\includegraphics[width=2.5cm]{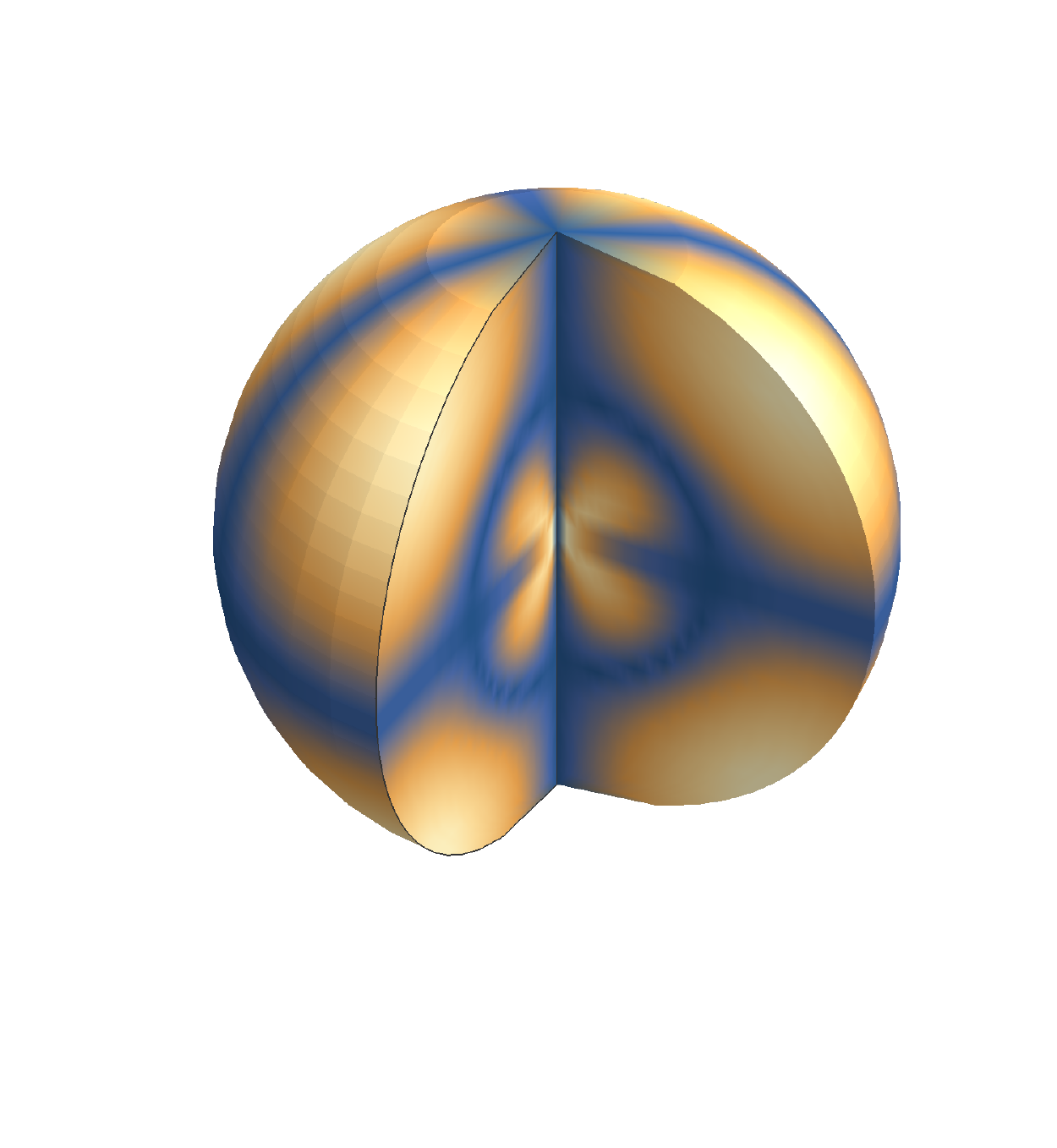}}
	\subfigure[(4+n)-$D$ $(x)$ mode]{\begin{minipage}[t]{0.12\textwidth}
		\includegraphics[width=2.5cm]{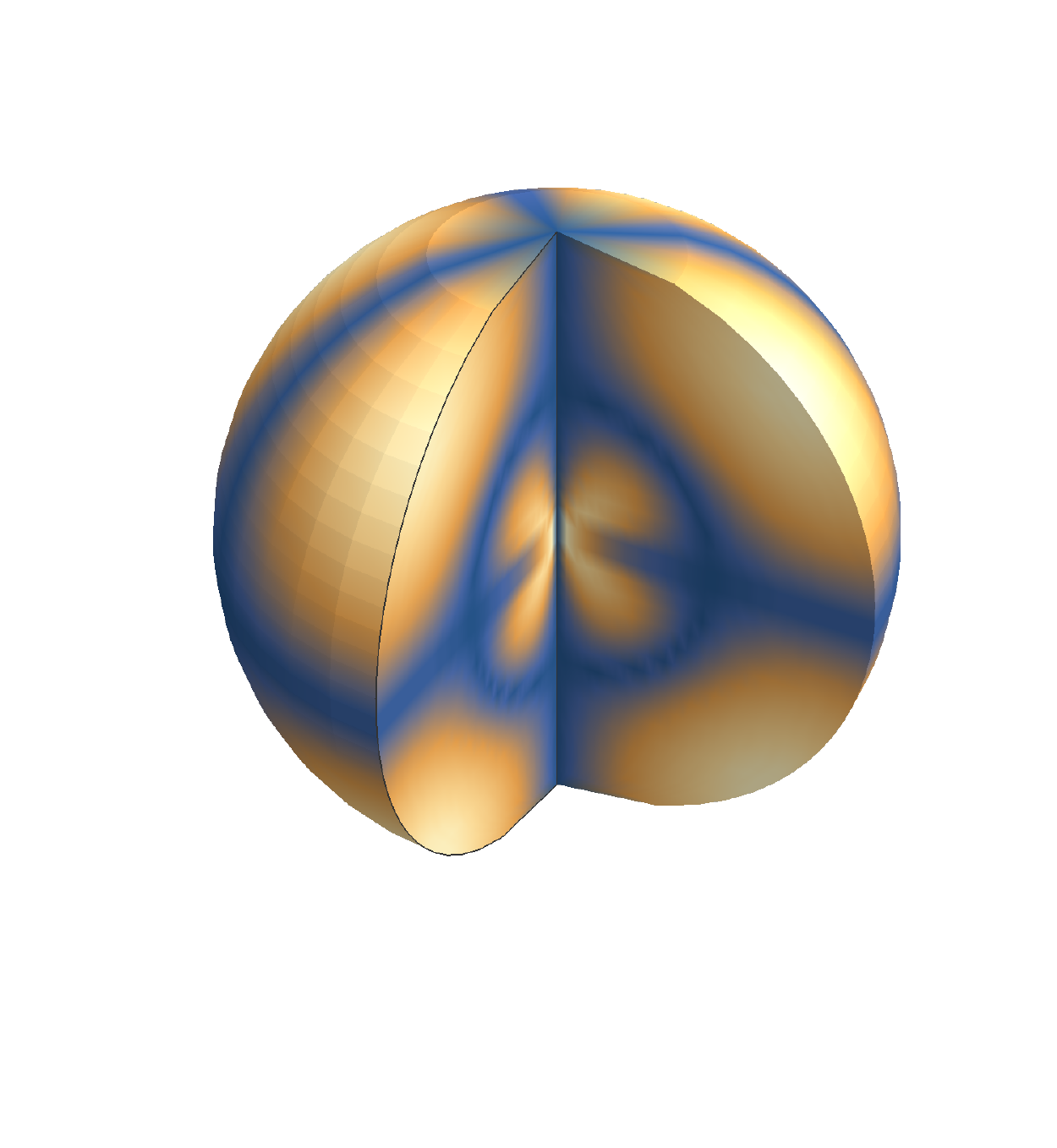}
	\end{minipage}
	\begin{minipage}[t]{0.12\textwidth}

		\includegraphics[width=2.5cm]{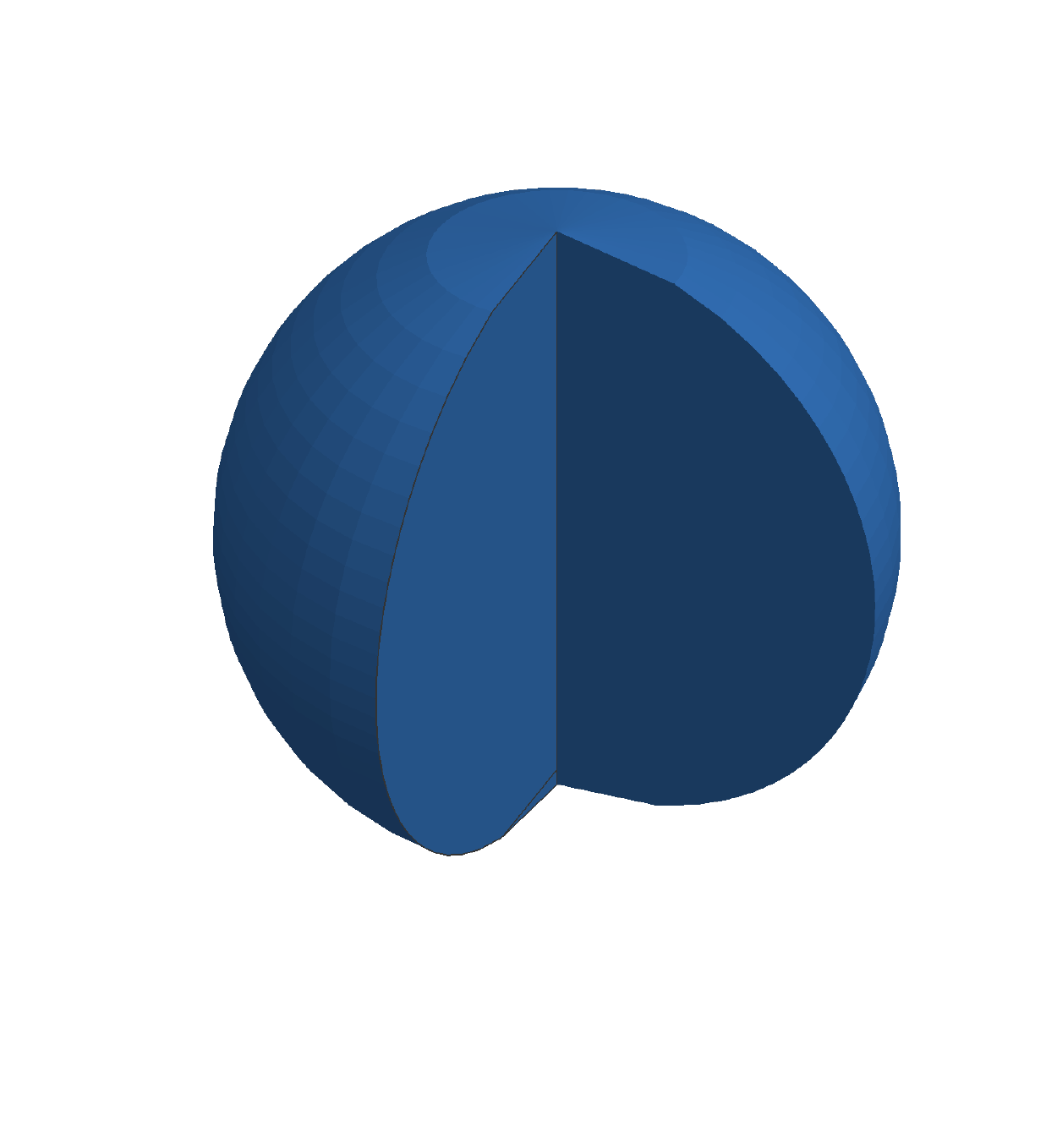}
	\end{minipage}}
	\\
	\subfigure[4-$D$ $(y)$ mode]{\includegraphics[width=2.5cm]{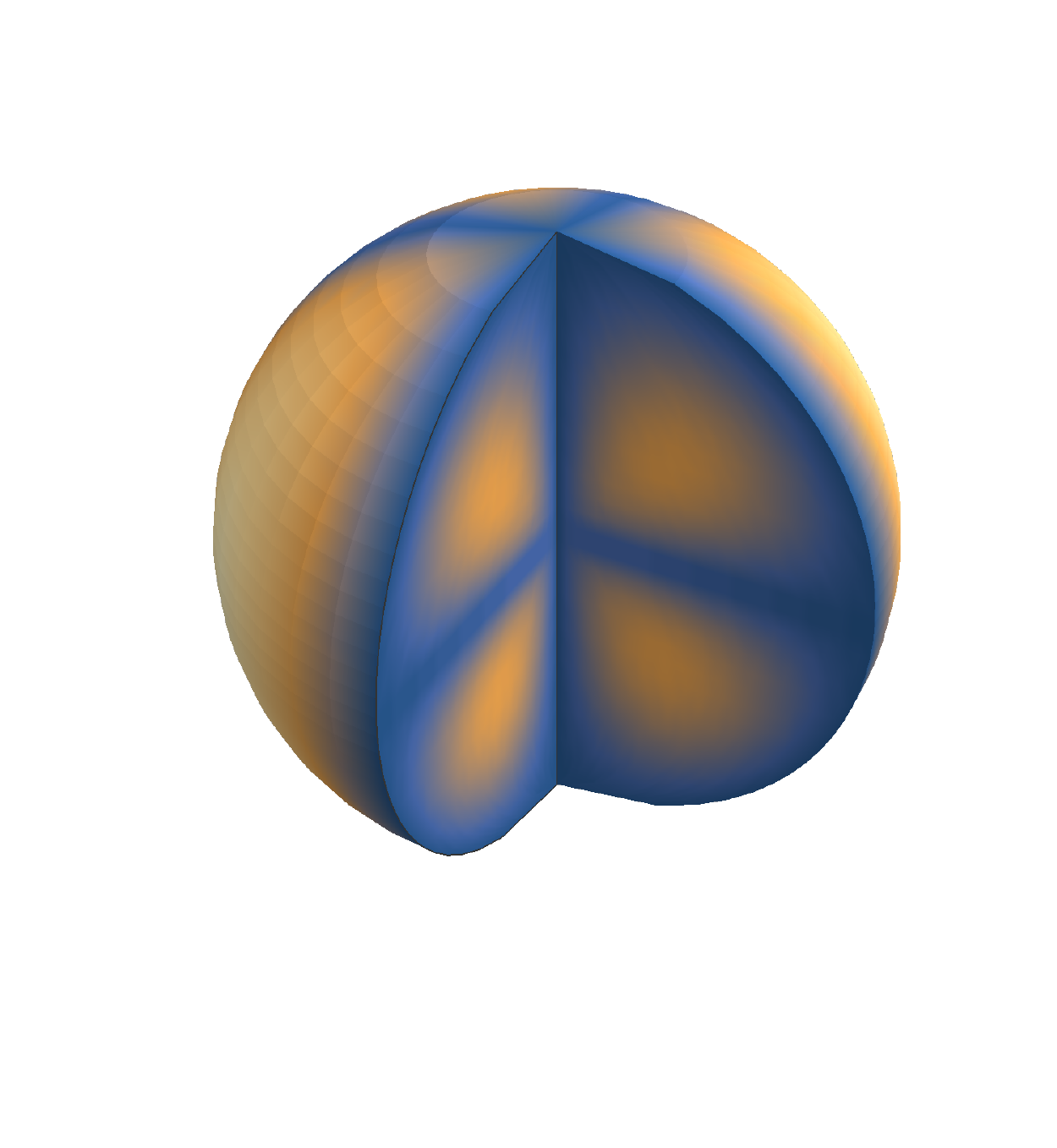}}
	\subfigure[(4+n)-$D$ $(y)$ mode]{\begin{minipage}[t]{0.12\textwidth}
		\includegraphics[width=2.5cm]{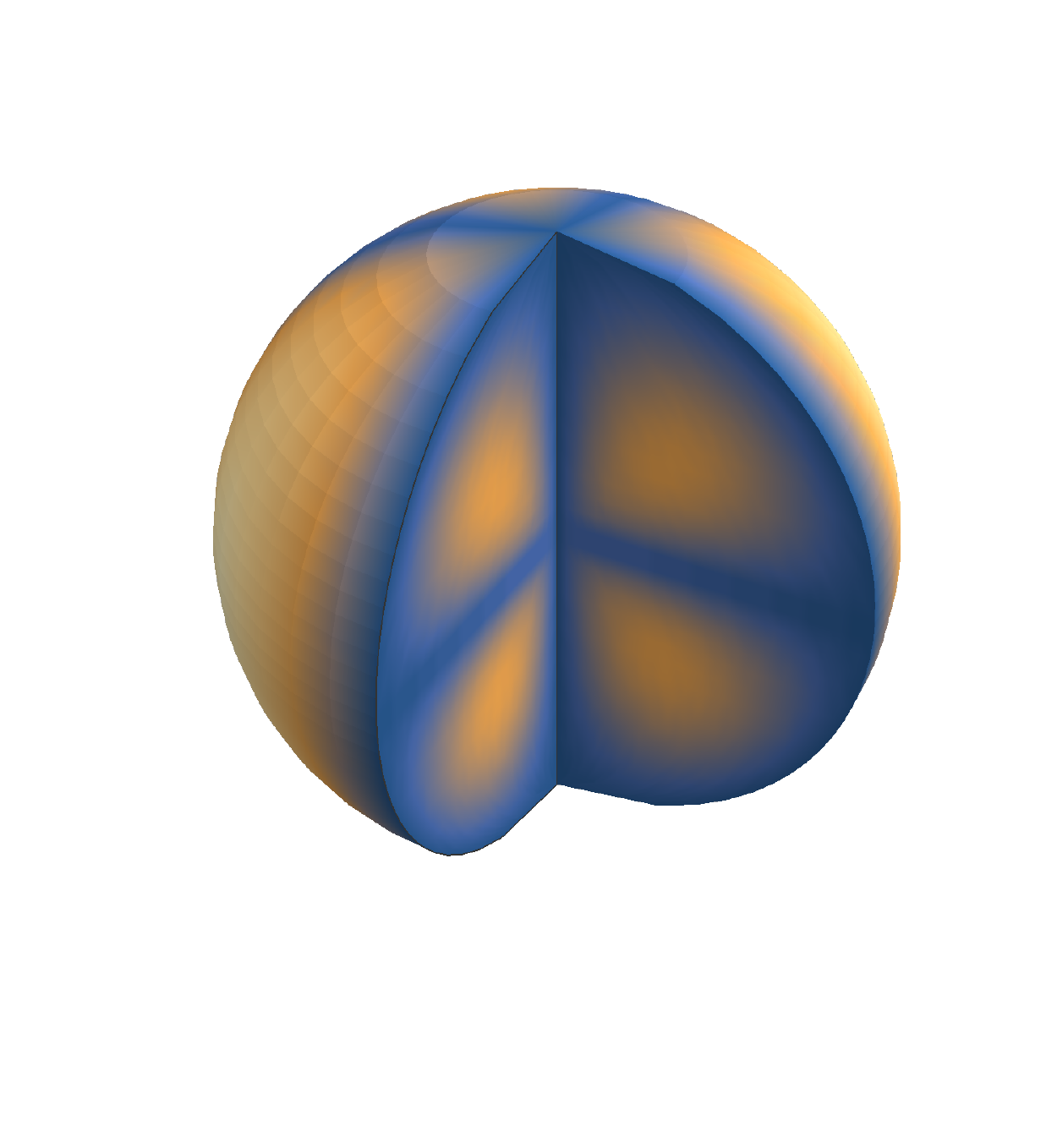}
	\end{minipage}
	\begin{minipage}[t]{0.12\textwidth}

		\includegraphics[width=2.5cm]{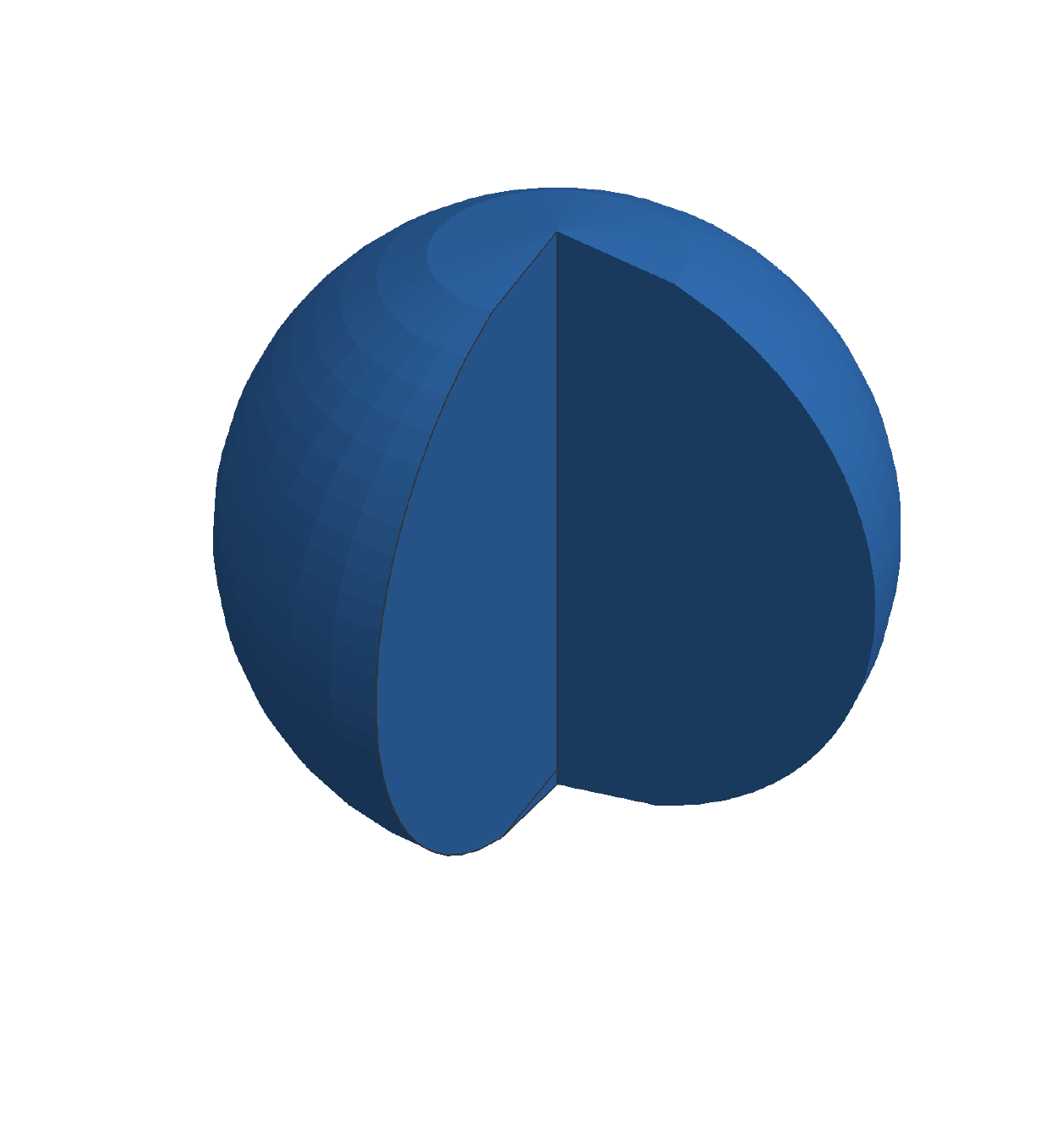}
	\end{minipage}}
	\caption{The responses of all the polarizations on the detector with (right column) and without (left column) extra dimensions.}
	\label{picture1}
\end{figure}

The set of pictures in FIG. \ref{picture1} illustrates with examples the response of each mode of gravitational waves in 3-dimensional subspace. In these 3-dimensional pictures, the vertical direction is the $z$-axis and the horizontal plane is the $(x, y)$ plane. We assume that the propagate directions of gravitational waves locate in 3-dimensional subspace, which is \textit{case 1} (For the general case, the situation gets more complicated, but it is still a linear combination of each mode according to Table \ref{table11}. ). The color of each point represents the response intensity of gravitational waves on the detector. The brighter the color, the stronger the response. And the direction of each point relative to the center of the unit ball represents the direction of the corresponding gravitational waves. But only know the direction is not enough to determine all the information of the propagation of gravitational waves, since waves can rotate in the spacelike 2-plane which perpendiculars to the spacial propagation direction in the 3-dimensional subspace. Thus, we set the radial coordinate to represent the angle of the rotation, whose range is $0\leq \alpha \leq \pi$.\par
 The pictures in the first column are the responses of the six polarizations without extra dimensions. Those in the second column are two possible responses of each polarization in the presence of one extra dimension $w$ (for the case of more extra dimensions, the results are similar). The polarization matrices of the gravitational waves of the second column are defined as follows. (1) We first consider gravitational waves propagating in the 3-dimensional subspace which we can observe. The 3-dimensional part of the polarization matrices of these gravitational waves are the same as that of the first column of the corresponding gravitational waves, and the remaining components of the polarization matrices are 0. (2) After that, we can obtain the polarization matrices of gravitational waves we need by taking a rotation in the ($x,w$) or ($y,w$) plane by an appropriate angle which will be given in detail later.
 If there are extra dimensions, although we assume that the direction of gravitational wave is still in the 3-dimensional subspace, the polarization matrices can have extra-dimensional components. And we can not see the response of these parts of the polarization matrices.
  \par
 In FIG. \ref{picture1}(a) and FIG. \ref{picture1}(b), the corresponding 3-dimensional part of the polarization matrices of the three pictures are
\begin{equation}
\label{E39}
	\left(
 	\begin{matrix}
 		h_{+} &0&0\\
 		0& -h_{+} &0\\
 		0&0&0\\
 	\end{matrix}
 	\right)
	, 
 	\left(
 	\begin{matrix}
 		0 &0&0\\
 		0& -h_{+} &0\\
 		0&0&0\\
 	\end{matrix}
 	\right),
 	\left(
 	\begin{matrix}
 		h_{+} &0&0\\
 		0& 0 &0\\
 		0&0&0\\
 	\end{matrix}
 	\right), 
\end{equation}
where the second and third matrices are obtained by taking a rotation of the first polarization matrix by 90 degrees in the ($x,w$) plane and ($y,w$) plane, respectively. For example, $p$ and $p'$ in FIG. \ref{p3} and Eq. (\ref{E34}). For all other possible cases, the 3-dimensional part of the polarization matrices can be expressed as a linear combination of the latter two matrices in Eq. (\ref{E39}). Therefore, the response of the detector in the 3-dimensional subspace in these cases can be represented by two pictures in FIG. \ref{picture1}(b). The same is true for all cases in the second column of FIG. \ref{picture1}.
 \par
 In FIG. \ref{picture1}(c) and FIG. \ref{picture1}(d), the corresponding 3-dimensional part of the polarization matrices of the left pictures are
\begin{equation}
	\left(
 	\begin{matrix}
 		0 &h_{\times}&0\\
 		h_{\times}& 0 &0\\
 		0&0&0\\
 	\end{matrix}
 	\right),
	\left(
 	\begin{matrix}
 		0 &h_{\times}&0\\
 		h_{\times}& 0 &0\\
 		0&0&0\\
 	\end{matrix}
 	\right),
 	\left(
 	\begin{matrix}
 		0 &0&0\\
 		0& 0 &0\\
 		0&0&0\\
 	\end{matrix}
 	\right),
\end{equation}
 where the second and third matrices are obtained by taking a rotation of the polarization matrix by 0 degrees and 90 degrees, respectively. The corresponding rotation plane is spanned by the $x$-axis (or $y$-axis) and any extra dimension. This means that in the presence of extra dimensions, we can see that the amplitude of the gravitational waves may become low or even zero, e.g. the response in the 3-dimensional subspace of the polarization matrix corresponding to $p'$ is lower than that of $p$. $p$ and $p'$ are two polarization planes showed in FIG. \ref{p3}.
\par
FIG. \ref{picture1}(f) and FIG. \ref{picture1}(h) describe the cases of breathing mode and longitudinal mode, respectively. The presence of the extra dimension does not have any effect on observations.\par
FIG. \ref{picture1}(j) and FIG. \ref{picture1}(l) describe the case of vector modes. The situation is the same as that in ($\times$) mode.\par

\par


\subsection{Distinguishing Features of Higher-dimensional and 4-dimensional Gravity Theories}
Different types of polarization modes could have fully correlated amplitude and same wavelength, phase and frequency in 3-dimensional subspace. This is one of the special and universal phenomena of gravitational waves propagating in higher-dimensional space-time. However, for modified gravitational theories in 4-dimensional space-time, different polarization modes are usually determined by different wave equations. And the waveforms of different polarization modes for same wave sources are usually different. For example, in case of a spherically symmetric core collapse, gravitational waves with tensor modes cannot be emitted, but those with breathing mode can be emitted \cite{Shibata:1994qd}. This means that different types of polarization modes respond differently to the symmetry of the source. This correspondence between polarization modes and source symmetry is fixed for 4-dimensional gravity theory, but no one has systematically studied it yet. However, for higher-dimensional gravity theory, since we can only see the projection of the higher-dimensional modes on the 4-dimensional brane, the correspondence between polarization modes and source symmetry is different from that in 4-dimensional gravity theory. For example, in Table \ref{table11}, the same mode in higher dimensions may project different modes on the 4-dimensional brane, but they respond in the same way to the same symmetry of the source. In addition, in many modified gravitational theories, the wave speeds of different modes of gravitational waves are also different. \par
 For example, for a higher-dimensional (+) mode, one may observe the (+) mode ($h_{+}$) and breathing mode ($h_{b}$) in the 3-dimensional subspace. See the polarization matrix of $p'$ in Eq. (\ref{E34}), although we can observe both tensor and breathing modes, since there is only one degree of freedom in higher dimensions, their phase, wavelength and frequency are the same, and their amplitude functions differ by only one coefficient, i.e. $h_{+}=a\times h_{b}$. That is to say, regardless of experimental precision, we should always observe this phenomenon if there are tensor modes in higher-dimensional gravitational waves, since mathematically the probability that the polarization plane is exactly at $p$ in FIG. \ref{p3} is zero. Thus in modified gravity theories in 4-dimensional space-time, gravitational waves of different polarization mode usually have different phase, wavelength, wave speed or waveform. Therefore, the observation of different polarization modes with the same phase, wavelength and frequency in 3-dimensional subspace may be one of the important evidence of extra dimensions. 

    Another unique phenomenon can be illustrated by tensor or vector mode in FIG. \ref{picture1}. In these examples, we have seen that by taking a rotation of the polarization matrices in the ($x,w$) or ($y,w$) plane, the amplitude of gravitational waves observed in the 3-dimensional subspace may decrease or even disappear. This shows that when we detect higher-dimensional gravitational waves in the presence of extra dimensions, the amplitude detected in the 3-dimensional subspace may be depressed or even vanishing. This is because we cannot detect all the information about the polarization of higher-dimensional gravitational waves in the 3-dimensional subspace. So the detected gravitational wave energy will be less than its total energy. This will cause the calculated luminosity distance of gravitational wave sources to be larger. If there is an electromagnetic counterpart corresponding to the gravitational waves, it will cause the calculated gravitational wave luminosity distance to be larger than the calculated electromagnetic wave luminosity distance. This phenomenon does not contradict the result of GW170817. It is worth noting that the above reason causing the gravitational wave amplitude to be lower is different from the way in the introduction (Gauss's law leads to faster decay of the Newton potential in higher-dimensional space-time).

   \par
  
 Here we only consider the theoretical analysis, so we ignore the accuracy of the detector. In practice, the signal will last for a period of time. During this period, due to the rotation of the earth or the detection satellite, there is usually a small change in the direction and spatial angle of the detector. These changes are smooth, which will cause the responses of the detector to be smooth curves on the unit sphere in pictures of FIG. \ref{picture1}. The current gravitational wave detectors adopt the method of matched filtering, which means that we can only verify whether there are known waveforms in the signals. Therefore, once we know the waveforms and the smooth curves caused by the detector movement, combined with FIG. \ref{picture1}, we can calculate the accurate waveform directly observed in the detector. If we can filter out these special waveforms, it may be an evidence of the existence of extra dimensions.\par

\subsection{What kind of extra-dimensional theory applies to our discussion?}
Previously, we only analyzed the case of flat extra dimensions, but a more general theory of extra dimensions requires considering warped extra dimensions. Here we consider a more general metric of extra dimensions theory
\begin{equation}
	ds^{2}=e^{2A(w)}\tilde{g}_{\mu\nu}(x)dx^{\mu}dx^{\nu}+g_{mn}(w)dw^{m}dw^{n},
\end{equation}
where $e^{2A(w)}$ is the warp factor and $g_{\mu\nu}(x,w)=e^{2A(w)}\tilde{g}_{\mu\nu}(x)$. This is the most general metric of extra dimensions that can guarantee the Poincare symmetry on the 4-dimensional brane.
Next, we will examine whether the classification in Table \ref{table11} can be generalized to curved extra dimensions. For \textit{case 1} in Table \ref{table11}, the tensor modes in higher dimensions can simulate the breathing modes on the brane. Now we will check whether this still holds in the case of warped extra dimensions. We consider the case where only tensor modes exist in the higher-dimensional spacetime. The simplest model in this case is the higher-dimensional general relativity, whose action has the following form
\begin{equation}
	S=\frac{1}{2\kappa_{D}}\int d^{D}x\sqrt{|g_{D}|}R_{D}.
\end{equation}
The linear perturbation of its equation of motion is
\begin{equation}
	\nabla^{p}\nabla_{p}h_{mn}-2R^{s}_{mnp}g^{pq}h_{qs}=0.
\end{equation}
For the sake of simplicity, we set $A=0$ and $g_{\mu\nu}=\eta_{\mu\nu}$. Then the Lorentz gauge can be written as
\begin{equation}
	g^{\pi\rho}\nabla_{\pi}h_{\rho\nu}-\frac{1}{2}\nabla_{\nu}h_{4}=\frac{1}{2}\nabla_{\nu}h_{N},
\end{equation}
where $h_{4}=g^{\mu\nu}h_{\mu\nu}$ and $h_{N}=g^{mn}h_{mn}$.
 The linear wave equations are
\begin{equation}
	\begin{split}
		\nabla^{\lambda}\nabla_{\lambda}h_{\mu\nu}=&0,\\
		\nabla^{\lambda}\nabla_{\lambda}h_{N}=&0.
	\end{split}
\end{equation}
When we consider the plane wave solution, we have
\begin{equation}
	\begin{split}
		h_{\mu\nu}=&e_{\mu\nu}e^{ik_{\rho}x^{\rho}},\\
		h_{N}=&f_{N}e^{ik_{\rho}x^{\rho}}.
	\end{split}
\end{equation}
Here we assume that the wave propagates along the $z$-axis direction. In addition to the four degrees of freedom constrained by the Lorentz gauge, we can also choose four constraints related to coordinate transformations. Here we choose $e_{0\nu}=0$. Then we have
\begin{equation}
	\begin{split}
		e_{0\nu}=e_{3\nu}=&0,\\
		e_{11}+e_{22}=&-f_{N}.
	\end{split}
\end{equation}
This means that, besides the two tensor modes, there is also a breathing mode simulated in this case. However, this is not the most general case of warped extra dimensions, but rather a Ricci-flat extra dimensions, due to our previous assumption ($A=0$, $g_{\mu\nu}=\eta_{\mu\nu}$ and $T_{mn}=0$ lead to Ricci flat extra dimension in this model).

\par

For \textit{case 2} in Table \ref{table11}, we use the method proposed in section 3.2 of Ref. \cite{andriot2017signatures}. This paper analyzes the case of compact warped extra dimensions by applied periodic boundary conditions. In order to extend their analysis to the case of infinitely large extra dimensions, we can simply let the period go to infinity.
 The conclusion is that the higher-dimensional tensor modes can simulate all other types of polarization modes in 4-dimensional space-time in this case. 

For the higher-dimensional breathing mode in \textit{case 2}, we can follow the same method as above and use this method in the Brans-Dicke theory which is the simplest theory that can produce breathing modes. A similar traceless condition can be proposed and this lead to the existence of non-zero longitudinal modes. That is, the longitudinal mode simulated by the breathing mode.
\par
Through the analysis of this subsection, we can generalize the conclusion of Table \ref{table11} to the general case of Ricci-flat warped extra dimensions.
\section{conclusion}
In this work, we mainly discussed the polarization modes of gravitational waves in higher-dimensional space-time and the possible observation in 3-dimensional subspace. In section \ref{section:2}, we defined the polarization modes of gravitational waves in higher-dimensional space-time.
 In section \ref{section:3}, we gave the transformation relations of the polarization matrices of the higher-dimensional gravitational waves under the Lorentz transformation, and classified the higher-dimensional gravitational waves according to these relations. Finally, in section \ref{section:4}, we analyzed the 3-dimensional observation effect of higher-dimensional gravitational waves. The contents include: (1) Which polarization modes will be observed in the 3-dimensional subspace by the gravitational waves of various polarization modes in the higher-dimensional space-time. (2) In the presence of extra dimensions, we found two unique phenomena when observing higher-dimensional gravitational waves in 3-dimensional subspace.

Of the two phenomena mentioned in the previous paragraph, the first is that for gravitational waves in higher-dimensional space-time, we may observe different polarization modes with the same phase, wavelength, frequency and waveform in 3-dimensional subspace. In other words, some polarization modes are simulated by other polarization modes through the projection. In the case of extra dimensions, this phenomenon is universal, and generally does not depend on a specific extra dimension theory. The second phenomenon is that in the presence of extra dimensions, since we can not measure all the information of polarization modes of higher-dimensional gravitational waves in the 3-dimensional subspace, our detection of gravitational wave amplitude and energy will be lower than the expected value, which will cause the measured luminosity distance to be larger than the real one. This is a unique phenomenon in extra dimensional models considered in this paper.\par

Note that we only considered the Minkowski background. Whether the space is compact or not will not change our results. These phenomena could be detected by space-borne gravitational wave detectors \cite{Lu:2019log} such as Lisa, Taiji and TianQin \cite{mei2021tianqin}.

 \section*{Acknowledgments}
We would like to thank Chun-Chun Zhu for useful discussion and Yi Chen for useful advice on English writing. This work is supported in part by the National Key Research and Development Program of China (Grant No. 2020YFC2201503), the National Natural Science Foundation of China (Grants No. 11875151, and No. 12047501), and the 111 Project (Grant No. B20063), Lanzhou City's scientific research funding subsidy to Lanzhou University, and  the Department of education of Gansu Province: Outstanding Graduate ``Innovation Star" Project (Grant No. 2022CXZX-059 and Grant No. 2023CXZX-057).\par

\appendix

\section{background}\label{appendix:2}
\subsection{Spinor formalism}
The mathematical foundations of spinor were first proposed by Cardan in 1913 \cite{cartan2012theory}, after that with the development of quantum mechanics in 1920s \cite{dirac1981principles}, it was discovered that spinor is essential to describe the intrinsic angular momentum. Spinor is the most basic representation of special unitary groups, thus we can apply a Lorentz transformation on it. It was until the 1960s that spinor was applied to classical gravity, and Penrose first found that 2-spinor representation can provide a very powerful method in the analysis of Einstein field equations in 4-dimensional space-time \cite{penrose1960spinor}. Inspired by this method, Newman and Penrose developed a useful tool named the Newman-Penrose(NP) formalism, which provides a null tetrad that can make the analysis of the equation of motion for a massless field easier \cite{newman1962approach}. Here we just list all the conclusions we need, for detail derivation, please refer to the original paper \cite{newman1962approach} and the book \cite{penrose1984spinors}.\par
According to Clifford algebra, the dimension of spinor space increases exponentially with the dimension $D$ of initial space-time, which is $2^{D/2}$. Therefore, the lower the dimension, the greater the advantage of the spinor formalism over the tensor formalism. In 4-dimensional space-time, the dimension of the reduced spinor space is 2, and this spinor space is complex and symplectic. Here we define the spin basis ($o,\iota$) of this space, whose inner products satisfy
\begin{equation}
\label{E32}
	[[o,\iota]]=1=-[[\iota,o]],[[o,o]]=[[\iota,\iota]]=0.
\end{equation}
Thus, the components of any spinor in this space with respect to the basis ($o,\iota$) are defined by
\begin{equation}
	\kappa=\kappa^{0}o+\kappa^{1}\iota.
\end{equation}
\par
The above spinor algebra is self-contained but cannot incorporate the world-vetors and world-tensors which are the common representation of general relativity. To overcome this problem, the complex conjugates of every elements in spinor space are necessarily involved. We denote complex conjugates by a bar following the tradition in Penrose's work \cite{penrose1960spinor}
\begin{equation}
	\overline{\kappa^{A}}=\bar{\kappa}^{A'},
\end{equation}
and all these new spinors form a new spinor space, which is the dual space of the original spinor space. Thus, the spinor representation of a world-tensor can be written as
\begin{equation}
	T^{a}_{bcd}=T^{AA'}_{BB'CC'DD'},
\end{equation}
which means that the tensor produce of a spinor and its complex conjugate represents a world-vector. We use capital letters ($A,B\cdots$) and lower case letters ($a,b\cdots$) to denote spinor indices and vector indices separately. This conclusion is not obvious, and there is a beautiful geometry proof on it in Penrose's book \cite{penrose1984spinors}. Then we can construct a null tetrad by this spin basis
\begin{equation}
	l^{a}=o^{A}\bar{o}^{A'},n^{a}=\iota^{A}\bar{\iota}^{A'},m^{a}=o^{A}\bar{\iota}^{A'},\bar{m}^{a}=\iota^{A}\bar{o}^{A'},
\end{equation} 
which is the famous NP formalism, and this null tetrad satisfies:
\begin{equation}
	l^{a}n_{a}=1,m^{a}\bar{m}_{a}=-1,
\end{equation}
while all the other scalar products vanish. 

\par
 All these conclusions have a solid mathematical foundation in the books of Penrose and Rindler \cite{penrose1984spinors,penrose1984spinors2}. 
\par
\par
\subsection{Geroch-Held-Penrose formalism}
The Geroch-Held-Penrose (GHP) formalism has been invented to study 4-dimensional space-time by admitting one or two preferred null directions, which is first proposed by Geroch, Held and Penrose \cite{geroch1973space}. And it has been generalized  to $D$-dimensional space-time by Durkee and collaborators in 2010 \cite{durkee2010generalization}. \par
In $D$-dimensional space-time, the basis we choose is 
\begin{equation}
	(l,n,m_{2},\cdots,m_{D-1}),
\end{equation}
where $l,n$ represent the unit vectors along null directions with the same spacial component $z$, $m_{i}$ represent the unit vectors along all the other spacial directions, where $i\in\lbrace 2, ..., D-1\rbrace$. $m_{i}$ are just same as all the spacial unit vectors except for $e_{z}$ in the Cartisian coordinate, thus we define the Cartisian coordinate as $(e_{t}, e_{z}, m_{i})$. The connection of the GHP coordinate and the Cartisian coordinate are $l=\frac{e_{t}+e_{z}}{\sqrt{2}}$, $n=\frac{e_{t}-e_{z}}{\sqrt{2}}$ and $m_{i}=m_{i}$. It is worth noting that, we usually call the components of the Weyl tensor and the GHP tensor in these formalisms the Weyl scalar and the GHP scalar \cite{d2022gravitational}. \par
Thus, we can define the GHP scalar $T$ with spin $s$ and boost weight $w$ if it transforms as 
\begin{equation}
	T_{i_{1}...i_{s}}\mapsto X_{i_{1}j_{1}}...X_{i_{s}j_{s}}T_{j_{1}...j_{s}},
\end{equation}
\begin{equation}
	T_{i_{1}...i_{s}}\mapsto \lambda^{w}T_{i_{1}...i_{s}},
\end{equation}
where $X\in SO(D-2)$, $i_{s}$ represent $m_{i}$ and $\lambda$ is the parameter of boost in the $(l, n)$ plane.
\par
These are all we need about the GHP formalism in this work.\par
\par

\section{Detailed calculation on the classification of nonnull gravitational waves}\label{appendix:1}
By doing the general Lorentz transformation (\ref{E42}) with the constraints (\ref{E44}) on the components of Riemann tensor, we can obtain
\begin{equation}
\label{E43}
	\begin{array}{lr}
	\begin{split}
		R_{lmlm}\mapsto & a^{4}R_{lmlm}+b^{4}R_{n\bar{m}n\bar{m}}+a^{3}bR_{lnlm}+ab^{3}R_{ln\bar{m}n}+\\&a^{2}b^{2}R_{lnln}-a^{2}b^{2}R_{lmn\bar{m}},
	\end{split}	\\
	\begin{split}
		R_{n\bar{m}n\bar{m}}\mapsto & d^{4}R_{lmlm}+c^{4}R_{n\bar{m}n\bar{m}}+d^{3}cR_{lnlm}+dc^{3}R_{ln\bar{m}n}+\\&d^{2}c^{2}R_{lnln}-d^{2}c^{2}R_{lmn\bar{m}},
	\end{split}	\\
	\begin{split}
		R_{lmmn}\mapsto & \frac{a^{2}\bar{d}^{2}}{2}R_{lmmn}+\frac{\bar{c}\bar{d}a^{2}}{2}R_{lnlm}+\frac{\bar{c}\bar{d}b^{2}}{2}R_{ln\bar{m}n}+\\&ab\bar{c}\bar{d}R_{lnln}+\frac{b^{2}\bar{d}^{2}}{2}R_{nmn\bar{m}}+\frac{a^{2}\bar{c}^{2}}{2}R_{lml\bar{m}},
	\end{split}	\\
	\begin{split}
		R_{ln\bar{m}n}\mapsto & bd^{3}R_{n\bar{m}n\bar{m}}+ac^{3}R_{lmlm}+\frac{c^{2}\bar{b}\bar{d}}{2}R_{lmmn}+\\&\frac{1}{2}[d^{2}(a\bar{a}d\bar{d}-c\bar{c}b\bar{b})+cbd^{2}]R_{ln\bar{m}n}+
		\\&\frac{1}{2}[c^{2}(a\bar{a}d\bar{d}-c\bar{c}b\bar{b})+adc^{2}]R_{lnlm}+
		\\& cd (a\bar{a}d\bar{d}-c\bar{c}b\bar{b})R_{lnln}+\frac{\bar{b}\bar{d}d^{2}}{2}R_{nmn\bar{m}}+\\&\frac{\bar{a}\bar{c}c^{2}}{2}R_{lml\bar{m}}-\frac{acd^{2}+bdc^{2}}{2}R_{lmn\bar{m}},
	\end{split}	\\
	\begin{split}
		R_{lnlm}\mapsto & db^{3}R_{n\bar{m}n\bar{m}}+ca^{3}R_{lmlm}+\frac{a^{2}\bar{b}\bar{d}}{2}R_{lmmn}+\\&\frac{1}{2}[b^{2}(a\bar{a}d\bar{d}-c\bar{c}b\bar{b})+adb^{2}]R_{ln\bar{m}n}+
		\\&\frac{1}{2}[a^{2}(a\bar{a}d\bar{d}-c\bar{c}b\bar{b})+cba^{2}]R_{lnlm}+
		\\& ab (a\bar{a}d\bar{d}-c\bar{c}b\bar{b})R_{lnln}+\frac{\bar{b}\bar{d}b^{2}}{2}R_{nmn\bar{m}}+\\&\frac{\bar{a}\bar{c}a^{2}}{2}R_{lml\bar{m}}-\frac{acb^{2}+bda^{2}}{2}R_{lmn\bar{m}},
	\end{split}	\\
\begin{split}
		R_{lnln}\mapsto & b^{2}d^{2}R_{n\bar{m}n\bar{m}}+a^{2}c^{2}R_{lmlm}+ac\bar{b}\bar{d}R_{lmmn}+\\&bd(a\bar{a}d\bar{d}-c\bar{c}b\bar{b})R_{ln\bar{m}n}+ac(a\bar{a}d\bar{d}-c\bar{c}b\bar{b})R_{ln\bar{m}n}\\&+(a\bar{a}d\bar{d}-c\bar{c}b\bar{b})^{2}R_{lnln}+b^{2}d^{2}R_{nmn\bar{m}}+ac\bar{a}\bar{c}R_{lml\bar{m}}\\&-\bar{a}\bar{b}\bar{c}\bar{d}R_{nml\bar{m}}-acbdR_{lmn\bar{m}},
	\end{split}	\\
	\begin{split}
		R_{nn}\mapsto & c^{2}\bar{d}^{2}R_{mm}+d^{2}\bar{c}\bar{d}R_{n\bar{m}}+cd\bar{c}^{2}R_{l\bar{m}}+cd\bar{d}^{2}R_{nm}+\\&c^{2}\bar{c}\bar{d}R_{lm}+c\bar{c}d\bar{d}R_{ln}+d^{2}\bar{d}^{2}R_{nn}+c^{2}\bar{c}^{2}R_{ll}+c\bar{c}d\bar{d}R_{m\bar{m}},
	\end{split}	\\
\begin{split}
		R_{ll}\mapsto & a^{2}\bar{b}^{2}R_{mm}+b^{2}\bar{a}\bar{b}R_{n\bar{m}}+ab\bar{a}^{2}R_{l\bar{m}}+ab\bar{b}^{2}R_{nm}+\\&a^{2}\bar{a}\bar{b}R_{lm}+a\bar{a}b\bar{b}R_{ln}+b^{2}\bar{b}^{2}R_{nn}+a^{2}\bar{a}^{2}R_{ll}+a\bar{a}b\bar{b}R_{m\bar{m}},
	\end{split}	\\
\begin{split}
		R_{m\bar{m}}\mapsto  & ac\bar{b}\bar{d}R_{mm}+\frac{1}{2}(bd\bar{a}\bar{d}+bd\bar{b}\bar{c})R_{n\bar{m}}+\frac{1}{2}(ac\bar{c}\bar{b}+\\&ac\bar{a}\bar{d})R_{l\bar{m}}+\frac{1}{2}(bc\bar{b}\bar{d}+ad\bar{b}\bar{d})R_{nm}+\frac{1}{2}(ac\bar{b}\bar{c}+\\&ac\bar{a}\bar{d})R_{lm}+\frac{1}{2}(ad\bar{b}\bar{c}+bc\bar{a}\bar{d})R_{ln}+bd\bar{b}\bar{d}R_{nn}+\\&ac\bar{a}\bar{c}R_{ll}+\frac{1}{2}(ad\bar{a}\bar{d}+bc\bar{b}\bar{c})R_{m\bar{m}}.
	\end{split}	\\

	\end{array}
\end{equation}
Consistent with the text, the terms contributing to tensor modes  are
\begin{equation}
	R_{lmlm}, R_{n\bar{m}n\bar{m}}, R_{lmmn},
\end{equation}
the terms contributing to vector modes are
\begin{equation}
	R_{lnlm},R_{ln\bar{m}n},
\end{equation}
the terms contributing to breathing mode are
\begin{equation}
	R_{m\bar{m}},R_{ll}, R_{nn},
\end{equation}
the term contributing to longitudinal mode is
\begin{equation}
	R_{lnln}.
\end{equation}
 \par
For $D$-dimensional space-time, it is obvious that our classification does not change with the increasing of the number of extra dimensions. Thus, we can obtain Eq. (\ref{E20}) by analogy with Eq. (\ref{E43}).
\bibliographystyle{unsrt}
\bibliography{reference1}

\end{document}